\documentstyle[12pt,aaspp4]{article}
\def\av     {$A_{\rm V}$}
\def\arcsec {$^{\prime \prime}$}
\def\arcmin {$^{\prime}$}
\def\chb    {c($H\beta$)}

\def\degr   {$^{\circ}$}
\def\etal   {{\it et~al.\/}}
\def\HI     {H~{\rm I}}
\def\HII    {H~{\rm {II}}}
\def\kms    {~km~s$^{-1}$}
\def\mo     {{$M_{\odot}$}}

\def\Rtoo    {$R_{23}$}

\input psfig

\begin{document}

\title{Chemical Properties of Star-Forming Emission Line Galaxies at 
	z=0.1 --- 0.5 }
\author{Henry A. Kobulnicky\footnote{Hubble Fellow} and
	Dennis Zaritsky}
\affil{University of California, Santa Cruz \\ Lick Observatory/Board
of Studies in Astronomy \\ 
Santa Cruz, CA, 95064  \\ 
Electronic Mail: chip@ucolick.org, dennis@ucolick.org}
\authoremail{chip@ucolick.org}
%\author{Refereed Draft of 26 July, 1998}
%\author{NOT FOR CIRCULATION}
\author{Accepted for Publication in {\it The Astrophysical Journal} }

\vskip 1.cm

\begin{abstract}

We measure the \HII--region oxygen and nitrogen abundances for 14
star-forming emission line galaxies (ELGs) at intermediate redshifts
($0.11<z<0.5$) using optical spectra obtained with the Keck II
telescope and LRIS multi-object spectrograph.  The target galaxies
exhibit a range of metallicities from slightly metal-poor like the LMC
(12+log(O/H)$\simeq$8.4) to super-solar (12+log(O/H)$\simeq$9.05) where
the solar value is 12+log(O/H)$\simeq$8.89.  Oxygen abundances of the
sample correlate strongly with rest--frame blue luminosities.  The
metallicity--luminosity relation based on these 14 objects is formally
indistinguishable from the one obeyed by galaxies in the local
universe, although there is marginal evidence (1.1$\sigma$) that the
sample is slightly more metal-deficient than local galaxies of the same
luminosity.  The observed galaxies exhibit smaller emission linewidths
than local galaxies of similar metallicity, but proper corrections for
inclination angle and other systematic effects are unknown.  For 8 of
the 14 objects we measure nitrogen-to-oxygen (N/O) ratios.  Seven of
the 8 systems show evidence for secondary nitrogen production, with
$log(N/O)>-1.4$ similar to local spiral galaxies.  These chemical
properties are inconsistent with unevolved objects undergoing a first
burst of star formation.  Comparison with local galaxies showing
similar chemical properties suggests that these intermediate$-z$
objects contain substantial old stellar populations which were
responsible for the bulk of the heavy element production presently seen
in the ionized gas.

Four of the 14 galaxies exhibit small half-light radii and narrow
emission line profiles (Compact Narrow Emission Line Galaxies---CNELGs;
Koo \etal\ 1995) consistent with small dynamical masses despite their
large optical luminosities and high levels of chemical enrichment.  We
find that the four CNELGs are indistinguishable from the 10 other
emission line galaxies (ELGs) in the sample on the basis of their
metallicity and luminosity alone.   Because of their morphological
similarity to \HII\ and spheroidal galaxies, CNELGs have been proposed
as the starburst progenitors of today's spheroidals.  Our assessment of
the stellar chemical abundances in nearby spheroidals reveals that the
majority of the CNELGs are presently $\sim$4 magnitudes brighter
and $\sim$0.5 dex more metal-rich than the bulk of the stars in well-known metal-poor
dwarf spheroidals such as NGC~205 and NGC~185.  Two of the four CNELGs
exhibit oxygen abundances higher than the planetary nebula oxygen
abundances in NGC~205, making an evolution between these two CNELGs and
metal-poor dwarf spheroidals highly improbable.   However, the data
are consistent with the hypothesis that more luminous and metal-rich
spheroidal galaxies like NGC~3605 may become the evolutionary endpoints
of some CNELGs after 1 to 3 magnitudes of fading.  We suggest that the
$z=0.1-0.4$ ELGs, and perhaps some of the CNELGs, are the precursors to
today's spiral galaxies during an episode of vigorous bulge star
formation $\sim$5 Gyr ago.

\end{abstract}

\keywords{ISM: abundances --- ISM: \HII\ regions --- 
galaxies: abundances --- 
galaxies: evolution ---
galaxies: starburst ---}

\section{Chemical Measurements from \HII\ Regions}

The evolution of galaxies is driven largely by the evolution of their
constituent stars.  Stars consume gas, provide luminous and mechanical
energy, and produce nearly all the elements heavier than helium in the
universe.  The chemical properties of stars and gas within a galaxy are
like a fossil record chronicling its history of star formation and its
present evolutionary status.  Emission lines from \HII\ regions have
long been the primary means of chemical diagnosis in local galaxies
(Aller 1942; Searle 1971; reviews by Peimbert 1975; Pagel 1986; Shields
1990), but this method has not yet been exploited for galaxies at
cosmological distances.  Even the largest HII regions at modest
redshifts require several-hour integration times on 10 m class
telescopes.  Nevertheless, tracking the chemical history of the
universe through \HII\ region abundances is a powerful, relatively
unexplored technique to probe the chemical content and ionization
conditions of the interstellar medium in the vicinity of vigorous star
formation activity.  Complications due to ionization uncertainties,
line width uncertainties, saturated lines, multiple velocity
components, and limited lines of sight that plague absorption
experiments are less severe or absent.  Since the star formation rate
appears to have been higher in the past (Madau \etal\ 1996; Lilly
\etal\ 1996) bright \HII\ regions should be plentiful at 
higher redshifts.

In this paper we present emission-line measurements of the chemical
properties of star--forming galaxies at $z=0.1$ to $z=0.5$.  Four of
the 14 observed galaxies, because of their small sizes and small
velocity widths, have been termed Compact Narrow Emission Line Galaxies
(CNELGs) by Koo \etal\ (1994) and Guzman \etal\ (1996).  
Because CNELGs and spheroidal galaxies have
similar half-light radii and velocity widths, Koo \etal\ propose the
intriguing hypothesis that CNELGs may evolve into the latter when the
starbursts fade.\footnote{We adopt the terminology proposed by Kormendy
\& Bender (1994) and Mateo (1994)  that diffuse elliptical-like systems
with central surface brightness $\geq$22 mag. arcsec$^{-1}$ and
$M_B\geq$-16 be called dwarf spheroidal galaxies.  These
objects are also termed dwarf ellipticals in the literature where
a consistent nomenclature has yet to emerge.} Here, we attempt to
test this evolutionary hypothesis and constrain the descendants of the
objects using the chemical properties of the star-forming gas combined
with the simple principle that the metallicity of a galaxy increases
monotonically with age.

This program is a also preliminary attempt to use the chemical
properties of star-forming objects to trace the metal-enrichment
history of the universe.  Observations of Lyman--$\alpha$  and metal
absorption lines at increasingly greater redshifts demonstrate that the
universe as a whole was once more gas rich and metal poor than the
typical galaxy observed today (Lanzetta, Wolf, \& Turnshek 1995; Lu
\etal\ 1997).   Individual galaxies, too, must have been more metal
poor in the distant past.  Observational programs, like this one,
involving a much larger sample of emission line objects at greater
redshifts will make it possible to sketch a chemical enrichment history
for the universe that is directly comparable to the results gleaned
from absorption line studies.

\section{Observational Program}
\subsection{Target Selection and Observations}

We selected target sources from the list of emission line objects
observed by Koo \etal\ (1995, 1997) and Guzman \etal\ (1996) with
redshifts from $z=0.10$ to $z=0.40$.  
They present spectroscopic data
showing that the targets exhibit emission line ratios consistent with
normal star-forming galaxies rather than AGN or Seyfert galaxies.  Our
selected sources lie within the fields SA-68, Lynx2 and the cluster
Cl0024+16 ($z=0.4$).  We included 4 CNELGs and 6 other
emission line galaxies (ELGs) with larger sizes and line widths.
The CNELGs have ID \#s 1--4 in Table~1 
and are characterized by half-light radii $<$2 kpc and emission
line widths $\leq$55 \kms.
Target names, positions, redshifts, lookback times\footnote{A cosmology
with $H_0$=50, $\Lambda=0$, and $q_0$=0.1 is assumed throughout.},
magnitudes, emission line velocity widths, and original references
appear in Table~1, along with an identifying numeral.  The 10 primary
targets lie in 4 different fields which we observed with the LRIS (Oke
1995) multi-object spectrograph on Keck II during 1997 November 30 and
December 1.  Each field required a different slitmask containing
between five and ten 1.5\arcsec\ wide slits of length $\geq$20\arcsec.
Slits were aligned parallel to the y-axis of the 2048x2048 CCD
detector.  The position angle of each mask was chosen to maximize the
number of available targets in the 8\arcmin\ LRIS field while
simultaneously retaining the full spectral coverage on the detector and
maintaining a slit orientation within 20\degr\ of the parallactic angle
to minimize light losses caused by differential atmospheric refraction
(Filippenko 1982).

Regions of the slitmask not devoted to primary targets are used to
obtain spectra of 17 additional confirmed galaxies in the Cl0024+16,
Lynx2, and SA-68 fields.  These secondary targets are selected
primarily by their fortuitous location on unused portions of the
detector.  If two or more objects occupied a convenient position, we
chose to observe the one with the bluest colors based on wide-band
photometry (SA-68, Lynx2---Munn \etal\ 1997; Cl0024+16---Koo 1998).
Four of the 17 objects show measurable [O~II] and [O~III] emission
lines.  We include them in the analysis that follows as a small
comparison sample of ``typical'' star-forming galaxies.  We refer to
them as the ``other field galaxies'' hereafter, although we have not
shown that they are characteristic  of field spirals or star-forming
objects.

We obtained between three and eight 20--minute exposures  of each
field through intermittent cirrus clouds.  Seeing averaged 1\arcsec\ --
1.5\arcsec.  Airmasses varied between 1.0 and 1.4.  A low-resolution
300 line/mm grating and 2048$^2$ CCD with 0.21\arcsec\ pixel$^{-1}$ in
the spatial direction and 2.4 \AA\ pixel$^{-1}$ in the dispersion
direction allowed us to cover a large wavelength range in excess of
4000\AA\ at a typical resolution of 9 \AA\ FWHM.  A different grating
tilt is used for each field to cover the wavelength range from the
redshifted [O~II] $\lambda$3727 to [O~III] $\lambda$5007.  When the
spectral range permitted, redshifted H$\alpha$ was also observed.  For
typical targets at a redshift of $z=0.4$, the observed wavelength range
runs from 5200~\AA\ to 9200 \AA.  Objects in each field lie at
different redshifts, so the wavelength coverage varied according to the
field observed and the position of each object within a given field.

\subsection{Calibration}

Images are bias subtracted and flat-fielded in the usual manner using
illuminated dome-flat frames.  We combine three or more similar images
to reject cosmic rays and improve signal-to-noise using an averaging
algorithm which first weighted each exposure by signal-to-noise in the
continuum part of the spectrum.  For Cl0024 fields, we reject several
low signal-to-noise exposures strongly affected by clouds.
Helium-argon arc lamps provide the wavelength calibration reference
for all masks except those in the Cl0024 field where night sky lines
are used due to the lack of suitable arc exposures.  Wavelength
calibration and slit-rectification is performed by the EXPECTOR
software (Kelson 1998).  The wavelength calibration is accurate to
$\sim$2 \AA\ in the red part of the spectrum and $\sim$5 \AA\ near
[O~II] $\lambda$3727 due to a  lack of strong reference lines in the
blue end of the arc calibration spectrum.  We correct for
illumination variations along each slit and variations from
slit-to-slit using exposures of the twilight sky.  To account for
wavelength--dependent atmospheric refraction, we fit a moderate order
polynomial to make a spatial trace of the galaxy's stellar continuum in
each exposure, and we constructed apertures for the extraction of
one--dimensional spectra.  One-dimensional spectra of each object are
extracted with an aperture of 10 to 12 pixels (2.1\arcsec\ --
2.3\arcsec).  The resulting 1-dimensional spectra, not corrected for
reddening, are shown in Figures~1a,b,c.  Where the dynamic range in a
spectrum is large, the spectrum is plotted a second time amplified by
an arbitrary scale factor for display purposes.  Redward of
8300~\AA\ the spectra become increasingly affected by noise due to
imperfectly subtracted night sky emission lines exacerbated by fringing
on the CCD detector.

Multiple observations of the spectrophotometric standard stars G193-74
and SA95-42 (Oke 1990) through several different slits are used to
construct a mean flux calibration for all of the masks and slits.  The
derived sensitivity solution has a pixel-to-pixel RMS of 4\% but has 
residual slopes of up to 15\% from the red to the blue end of the
spectrum.  We believe most of the uncertainty on the slope of the
sensitivity calibration is caused by light losses on either side of the
slit due to differential atmospheric refraction during standard star
exposures.  The impact of this effect is small and is discussed 
below along with uncertainty estimates.

\section{Spectral Analysis}

\subsection{Line Ratios and Reddening}

The flux in each emission line is measured using a single Gaussian fit,
taking the width, height, and position of the Gaussian as free
parameters.  For lines with low signal-to-noise, the width and position
of the Gaussian is constrained using the redshift and line width
computed from other nearby stronger emission lines.  In cases where a
line is not visible at the expected place in the spectrum, the
3$\sigma$ upper limit is calculated from the pixel-to-pixel RMS at that
location and the square root of the number of pixels across a typical
line, i.e., 3$\times$RMS$\times{\sqrt{9}}$.

Observed Balmer line ratios allow us to deredden each emission line
relative to H$\beta$ using the nebular emission line analysis software
described in Kobulnicky \& Skillman (1996) based on the standard
nebular analysis techniques (e.g., Osterbrock 1989).  All observed
fields lie at high Galactic latitude, and reddening due to dust within
the Milky Way produces extinctions of \av$<$0.11 mag.  corresponding to
c(H$\beta$)$<$0.05 (Burstein \& Heiles 1984; Schlegel, Finkbiner, \&
Davis 1998).  Thus, we assume that all of the extinction occurs within
the target galaxy and we deredden the spectra in the rest frame of the
source using the Galactic extinction law (Seaton 1979 as parameterized
by Howarth 1983).  Use of an LMC and SMC extinction curve would not
significantly alter the results since they do not differ appreciably
from the Galactic one for the optical wavelengths of interest here.

Logarithmic extinction parameters, c(H$\beta$), underlying stellar
hydrogen absorption, EW(abs), and the electron temperatures are derived
iteratively and self-consistently for each object.  The theoretical
Balmer line ratios (Hummer \& Storey 1987) are not strongly dependent
on the electron temperature of the nebula, so an initial guess of
10,000 K produced reddening estimates not far from the final best-fit
values.  For spectra where H$\alpha$, H$\beta$, and H$\delta$ are all
measured, we solve for the combination of c(H$\beta$), and EW(abs)
which best match the theoretical Balmer line ratios.  One object,
L2-408115 shows significant evidence for Balmer absorption, while SDG~231
and SDG~146 have marginal indications of stellar absorption.  When only
H$\beta$ and H$\delta$ are measured, EW(abs) is fixed at 2 \AA\ which
is representative of local irregular, \HII\, and spiral galaxies,
(McCall, Rybski, \& Shields 1985; Izotov, Thuan, \& Lipovetsky 1994).
Three objects which have only 1 measured Balmer line are assigned
EW(abs)$\simeq$2 \AA\ and c(H$\beta$)=0.0$\pm$0.15.  Tables~2---3
contain dereddened line fluxes relative to H$\beta$ for each observed
target using

\begin{equation}
{{I(\lambda)}\over{I(H\beta)} } = { {F(\lambda)}\over{F(H\beta)} }
10^{c(H\beta)f(\lambda)}  
\end{equation}

\noindent where $I$ is the true de--reddened flux at a given
wavelength, $F$ is the observed flux at each wavelength, c(H$\beta$) is
the logarithmic reddening parameter which describes the amount of
reddening relative to H$\beta$,  and f($\lambda$) is the
wavelength--dependent reddening function which is listed for each
line.

\subsection{Measurement Uncertainties}

Due to intermittent cirrus clouds, uncertainties on the zero point of
the flux calibration may be as large as a factor of 2 based on the
$\sim$1 mag. variation among standard star exposures.   Absolute
quantities such as the H$\beta$ flux and the derived star formation
rates will carry this uncertainty while measurements requiring only the
ratios of emission lines are much more robust, as discussed below.  We
observed standard stars in relatively good sky conditions so that the
H$\beta$ flux, and the derived star formation rates should represent a
lower limit on their true values.

Uncertainties on the emission line ratios are dominated by three
factors:  1) the quality of the sky subtraction, especially in spectral
regions affected by strong night sky OH emission lines, 2) the quality
of the flux calibration, and 3) the accuracy of the reddening
determination.  Most of the objects lie at redshifts where the common
nebular emission lines lie in close proximity to strong night sky
lines.  For this reason, we adopt an empirical method of error
estimation for each emission line using the RMS in the adjacent
continuum portion of the spectrum.  The 1$\sigma$ uncertainties are
computed from the offline RMS and are added in quadrature to
uncertainties on the flux calibration and reddening correction as
described below.  The formal pixel-to-pixel RMS in the sensitivity
curve is 0.04.  However, sensitivity solutions using different standard
stars exhibited relative slopes of up 15\% from the red to blue end of
the spectrum.

For purposes of obtaining accurate emission line ratios needed here, it
is possible to compensate for a linear uncertainty in the slope of the
flux calibration by application of a reddening correction determined
from Balmer line ratios.  The derived value of the reddening parameter,
\chb, will be systematically incorrect, but its application to the data
will compensate, to first order, for a linear error in the sensitivity
solution.  With this in mind, we solved for the best-fit value of
\chb\ and record it in Tables~2--3 for each object.  The typical
uncertainty on \chb\ is 0.05 where H$\alpha$ and H$\beta$ are
well-measured.  This error is added in quadrature with the 4\%
uncertainty on the flux calibration and the empirically-determined
uncertainties due to photon, sky, and detector noise to produce the
final 1$\sigma$ uncertainties in Tables~2--3.  For objects without an
H$\alpha$ detection, the reddening correction becomes more uncertain
since the higher order Balmer lines are weaker and are affected by
underlying stellar absorption.  In these cases we use the best-fit
values determined from H$\beta$ and H$\gamma$.  We assign an
uncertainty of $\delta$\chb=0.10 which gets propagated into the final
tabulated uncertainties.  An error of $\delta$\chb=0.10 is equivalent
to a 15\% uncertainty in the slope of the flux calibration over the
optical wavelength range when considering the total error budget.  In
the three cases where H$\beta$ is the only Balmer line detected, the
reddening is essentially unconstrained, and we assign
$\delta$\chb=0.15.  Removal of these three objects 
from the sample would not alter the conclusions 
of the subsequent sections.  

\section{Assessing the Physical and Chemical Properties of the Sample}

\subsection{Direct Oxygen Abundance Measurements} Electron temperature
and heavy element abundance strongly influence the nebular spectrum of
a star-forming region.  The ratio of a high-excitation auroral line
such as [O~III] $\lambda$4363 to the lower excitation [O~III]
$\lambda\lambda$4959,5007 lines provides a direct measurement of the
electron temperature in the physical medium where $O^{++}$ is the
dominant ionization state.  In metal-deficient \HII\ regions where the
relative paucity of heavy elements reduces the cooling capability of
the ISM, the upper levels of $O^{++}$ are strongly populated and
[O~III] $\lambda$4363 is prominent in the spectrum.  In relatively
metal-rich environments such as the Galaxy and even the LMC, the
temperature sensitive [O~III] $\lambda$4363 line is seldom detected
since the enhanced cooling due to abundant heavy elements reduces
collisional excitation of the upper levels.

Three objects in the sample exhibit measurable  [O~III] $\lambda$4363
which enables a direct determination of the electron temperature.  For
L2-10083, L2-11500, and SDG~223, $T_e([O~III])$ is 16100$\pm$500 K,
11000$\pm$500 K, and 10100$\pm$400 K.  We determine $T_e([O~II])$ of
the lower ionization zones using the empirical fit to photoionization
models from Pagel \etal\ (1992) and Skillman \& Kennicutt (1993),

\begin{equation} T_e(O^+)=2(T_e^{-1}(O^{++}) + 8\times10^{-5})^{-1}.
\end{equation}

For the majority of objects, only upper limits on $T_e([O~III])$ can be
measured.  In most cases, the upper limits do not place strong
constraints on the electron temperature.  Table~4 lists T$_e$ as
determined for each object determined from the direct detection of
[O~III] $\lambda$4363 or an upper limit on [O~III] $\lambda$4363.  The
next column lists the oxygen abundance derived from the given
temperature and [O~II] and [O~III] lines the assuming that

\begin{equation}
{ O\over H} = {{O^{+}}\over{H^{+}}} + {{O^{++}}\over{H^{+}}} .  
\end{equation}

\noindent 
For L2-10083, L2-11500, and SDG~223, the computed values of 12+log(O/H)
are 7.84$\pm$0.04, 8.22$\pm$0.04 and 8.28$\pm$0.04.    We discuss the
other derived parameters in \S~4.2 and \S~4.3.

There are several reasons why the electron temperatures derived for
these objects may be regarded as {\it upper limits}, and the resulting
oxygen abundances taken as {\it lower limits} to the actual values.
Emission lines with low S/N, such as the [O~III] $\lambda$4363 lines
measured here, tend to be  systematically overestimated by the Gaussian
fitting technique in the presence of significant noise.  Furthermore,
the presence of temperature variations within the measured objects will
bias the spectrum toward the highest temperature, highest emissivity
regions.  This effect has long been known to produce overestimates of
the electron temperatures within individual \HII\ regions (Peimbert
1967).  This same effect appears to bias the integrated spectra of
nearby star--forming irregular and blue compact galaxies toward larger
electron temperatures compared to spatially-resolved spectra of
individual \HII\ regions and diffuse gas (Kobulnicky, Kennicutt \&
Pizagno 1998).  Because these two systematic effects bias the computed
oxygen abundance toward lower values, we regard the O/H estimates from
the direct method as firm lower limits.

\subsection{Empirical Oxygen Abundances}

Oxygen abundances may also be determined from the ratio of
[O~II]+[O~III] to H$\beta$ lines even when the electron temperature is
not measured directly.  Because the relationship between oxygen
abundance and $R_{23}$ ($\equiv(I_{3727} +I_{4959}+I_{5007})/H\beta)$)
is double valued, this so-called ``strong-line'' method is useful only
when the approximate metallicity is known a priori.  Multiple authors
have developed formulae for converting $R_{23}$ into oxygen abundance
both for very metal poor (12+log(O/H)$<$7.9---Pagel, Edmunds \& Smith
1980; Skillman 1989) and metal-rich regimes (12+log(O/H)$>$8.3---Pagel
\etal\ 1979; Edmunds \& Pagel 1984; McCall, Rybski, \& Shields 1985;
Dopita \& Evans 1986).  We adopt the analytical approximation for the
metal-rich branch of Zaritsky, Kennicutt, \& Huchra (ZKH, 1994) which
is itself a polynomial fit to the average of three earlier calibrations for
metal-rich \HII\ regions.  

Figure~2 shows each of these previous calibrations of \Rtoo\ along with
the empirical fit of ZKH.  Figure~2 also displays the locations of the
14 emission line objects along the \Rtoo--O/H relation.  Symbols
distinguish the four designated CNELGs from the six other ELGs.  Filled
triangles denote the four field galaxies with measurable emission
lines.  Figure~2 reveals that the target galaxies exhibit a wide range
of oxygen abundances from 12+log(O/H)=9.05 (1.45 the solar value) to
12+log(O/H)=8.25 (0.23 the solar value).  The $R_{23}$ indicator is
especially uncertain in the turn-around region near 12+log(O/H)=8.3.
Fortunately,  only L2-10083 falls into this turn-around region, and it
has a direct abundance estimate which provides a firm lower limit to
its oxygen abundance.

Table~4 lists the oxygen abundances computed from the ZKH empirical
relation for each object.  The quoted errors are uncertainties due to
measurement errors only and do not take into account systematic effects
in the calibration. In the subsequent analysis, we add an additional
error term of 0.2 dex in quadrature with the uncertainties due to
observational effects to obtain a realistic estimate of the total
uncertainty on the oxygen abundance.  The solid lines in Figure~2
representing previous \Rtoo\ calibrations provide an indication of the
magnitude of systematic effects.  In nearly all cases, the
observational uncertainties ($\sim$0.1 dex) are smaller than the
potential systematic uncertainties ($\sim$0.15--0.2 dex).  We also
provide in Table~4 an ``indicative electron temperature'', $T_{e-ind}$,
which is the temperature that would yield the same O/H ratio determined
from the empirical calibration.  We adopt uncertainties on $T_{e-ind}$
which produce O/H uncertainties that match the uncertainties due to the
measurement errors in the empirical calibration.

It is instructive to compare the results obtained from the direct
abundance determination with the strong-line approach.  In Figure~2,
dotted lines connect the oxygen abundances derived using the two
different methods.  The strong-line method yields an oxygen abundance
which is systematically higher by 0.2--0.3 dex for the three objects
where comparison is possible.  As discussed above, the direct O/H
determinations should be regarded as lower limits given the biases in
the measured electron temperature.  However, these three direct
abundance measurements serve an important role because they constrain
the oxygen abundances to lie on the upper, metal-rich branch of the
\Rtoo\--O/H calibration.  Three additional objects (SA68-206134, SDG~125,
and SDG~183) have $T_e$ upper limits which constrain the oxygen
abundance to lie on the metal-rich branch as well.  Given that these
six well-constrained objects all lie on the metal-rich branch of
\Rtoo\ calibration, it is probable that the remaining 8 objects also
lie on the metal-rich branch.  If there exist no biases that would
cause us preferentially to measure electron temperatures in objects on
the metal-rich branch of the \Rtoo\ calibration,\footnote{Actually, the
converse is more probably true.  Low abundance ($Z<0.3 Z_\odot$)
objects have higher
electron temperatures and stronger  [O~III] $\lambda$4363 emission
lines than objects near solar metallicity.  Objects in the 
abundance abundance range $0.05 Z_\odot< Z <0.3 Z_\odot$ will
have the strongest [O~III] $\lambda$4363 lines and have
the most easily-measured electron temperatures.} then, by the binomial
theorem, the observation that 6 out of 14 objects are on the metal-rich
branch allows us to reject the possibility that the sample contains 1
or more objects on the metal-poor branch at the 43\% confidence level.
The probability that 2 of the remaining objects are actually on the
low-metallicity branch is 30\% or less.  There is only a 0.03\%
probability that {\it all} of the remaining 8 objects could actually be
low-metallicity objects for which use of the metal-rich
\Rtoo\ calibration would be inappropriate.  Thus, we adopt the ZKH
\Rtoo\ metallicity calibration for all of the targets studied here.

\subsection{Nitrogen Abundances}

Seven of the sources exhibit measurable [N~II] $\lambda$6584 lines.
Since longer-lived intermediate mass (4--8 \mo) stars produce much of
the N in galaxies (Renzini \& Voli 1981), measurement of the N/O ratio
provides information about the existence and nucleosynthetic
contribution of stars in this mass range.  We use [N~II] $\lambda$6584
in conjunction with [O~II] $\lambda$3727 and $T_{e-ind}$  to make an
estimate of the N/O ratio assuming,

\begin{equation}
{ N\over O} = {{N^{+}}\over{O^{+}}}  
\end{equation}

\noindent Equation~3 appears justified in low metallicity environments
where photoionization models indicate uncertainties of less than 20\%
result through this approximation.  However, in environments near solar
metallicity, this approximation may underestimate the true N/O ratios
depending on the nature of the ionizing spectrum (Garnett 1990).  Since
the excitation energies of the two collisionally-excited lines are not
vastly dissimilar, the resulting N/O ratio is not strongly dependent on
the adopted electron temperature.  Table~4 lists the N/O ratios for the
7 objects with [N~II] $\lambda$6584 detections and N/O upper limits for
the two objects in which an upper limit on [N~II] $\lambda$6584 is
measured.  The seven N/O measurements range between log(N/O)=$-1.08$
and log(N/O)=$-1.59$.  Uncertainties due to reddening and sky
subtraction in the presence of strong night sky emission lines near
[N~II] $\lambda$6584 dominate the error budget for N/O.

Figure~3 displays the oxygen abundances (12+log(O/H)) and nitrogen
abundances (log(N/O)) for the observed objects.   Symbols differentiate
the abundances of CNELGs (filled circles), ELGs  (open circles), and
the field galaxies with emission lines (filled triangle), nearby dwarf
irregular and blue compact galaxies (open squares; from the compilation
of Kobulnicky \& Skillman 1996), the LMC (filled square; error bars
represent the dispersion among LMC \HII\ regions; Pagel \etal\ 1978),
and low-surface-brightness galaxies (stars; van Zee 1998).  With the
exception of L2-410083, all of the observed objects appear chemically
similar to local spiral galaxies.  They exhibit oxygen abundances
higher than those observed in the LMC.  Their mean N/O ratios are
log(N/O)$\sim-1.25$, consistent with nitrogen from secondary\footnote{
A ``primary'' nucleosynthesis product is generally used to mean an
element generated directly from the initial H and He present in the
star, while a ``secondary'' nucleosyntheis element is one produced from
pre-existing heavy elements through processes such CNO
cycling.} nucleosynthesis processes which apparently produce much of
the nitrogen in local spiral galaxies (e.g.,  Vila-Costas \& Edmunds
1993 and references therein).

\subsection{On the Suitability of Integrated Galaxy Spectra
for Chemical Analysis}

The physical properties of ionized gas in galaxies vary considerably on
both small and large scales.  Spiral galaxies, especially those lacking
visible bars, exhibit considerable radial chemical gradients, sometimes
spanning more than a dex in O/H from nucleus to outer disk (e.~g.
Zaritsky, Kennicutt \& Huchra 1994;  Martin \& Roy 1994).  The spectral
shape of the ionizing radiation field varies with location in
star--forming galaxies, producing both concentrations of high
ionization gas in \HII\ regions and extended low-ionization filaments
which sometimes stretch for a kpc or more (e.g., Hunter 1994; Hunter \&
Gallagher 1997).  When viewed from large distances or with low spatial
resolution, a galactic spectrum will contain an ad hoc mixture of high
and low-metallicity gas with high and low ionization characteristics.
Is a spatially-integrated galaxy spectrum in any way a meaningful
measure of the physical conditions contained within?

In the case of low-mass irregular and blue compact galaxies there
appear to be no significant chemical variation (Kobulnicky \& Skillman
1997).  Except for occasional biases that lead to small overestimates
of the mean electron temperature, the presence of varying ionization
conditions throughout such systems does not preclude meaningful chemical
measurements.   Furthermore, such galaxies are often dominated by a
single giant \HII\ region which contributes most of the light to the
integrated spectrum.  Integrated spectra yield direct measurements of
the oxygen abundances which are in good agreement with localized
measurements in individual \HII\ regions (Kobulnicky, Kennicutt, \&
Pizagno 1998).  Even in larger spiral galaxies where multiple
\HII\ regions exhibit substantially different chemical properties and
ionization levels, the oxygen abundances measured from the strong-line
ratio method produce mean abundance estimates that correlate strongly
with the mean abundances at 0.4 isophotal radii.

The objects in this study span a range of velocity dispersion
($\sigma_v=$29 \kms\ --- 200 \kms), magnitudes ($M_B=-$16.6 --- 22.0)
and blue half-light radii (1.2 kpc --- 6.7 kpc).  By comparison with
local galaxies, the velocity widths and luminosities of line-emitting
regions in the smallest, least luminous targets are consistent with a
giant \HII\ region ($\sim$1 kpc) such as 30 Doradus or NGC~604
(Terlevich \& Melnick 1981).  The ELGs reported by Guzman \etal\ (1998)
appear to have a high surface brightness residual cores of diameter
$\sim1\ kpc$ after subtraction of an exponential disk.  This size is
consistent with giant \HII\ complexes in local galaxies which appear
chemically homogeneous (Kobulnicky \& Skillman 1997).  Thus, in these
smallest objects, the integrated spectrum is dominated by emission over
a relatively small region, and is likely to be chemically
representative of the system as a whole.

In the larger ELGs, the linewidths are larger than in local giant
\HII\ regions  and the line-emitting region is probably spread over an
area of the host galaxy larger than a few kpc.  Point-to-point chemical
variations may be present at the level of 0.2 dex as in the LMC (Pagel
\etal\ 1979).  In the most massive galaxies, SDG~125 and SDG~146, the
emission line profiles show evidence of considerable rotational
broadening and spatial chemical variations are probably present,
perhaps up to 1 dex if they are similar to local spirals.
Unfortunately the spatial resolution of our ground-based slit spectra
is too poor to constrain the size of the nebular region.

During the analysis that follows, we will adopt the motivated
assumption that that the emission line spectra of all objects are at
least roughly representative of the chemical content of the galaxy as a
whole, and certainly not gross overestimates.

\subsection{Emission Line Luminosities and Star Formation Rates}

Table~5 lists the H$\beta$ luminosities in the source rest frame
assuming $H_0$=50 \kms\ Mpc$^{-1}$, $q_0=0.1$ and $\Lambda=0$.  We also
list the H$\beta$ luminosity corrected for extinction within the target
galaxy, and the mass of ionized hydrogen, using emissivities tabulated
in Osterbrock (1989) for Case B recombination, assuming $T_e=10,000~K$,
and $n_e=100$ electrons cm$^{-3}$.  The H$\beta$ luminosities range
from 0.4$\times10^{41}$ erg s$^{-1}$ to 40$\times10^{41}$ erg s$^{-1}$
corresponding to log($L_{H\alpha}$) from 41.0 to 43.0 (assuming
$L_{H\alpha}/L_{H\beta}=2.86$ for typical \HII\ regions).  These
emission line luminosities span the upper end of the range of normal
spiral galaxies which have $40.8<log(L_{H\alpha})< 42.5$ (Kennicutt
1983).  Adopting

\begin{equation}
SFR(total) = {{L(H\alpha)}\over{1.12\times10^{41}} erg~s^{-1} } M_\odot
~yr^{-1}~~(Kennicutt~1983), 
\end{equation}

\noindent the derived star formation rates range from 0.36
$M_\odot~yr^{-1}$ to 35  $M_\odot~yr^{-1}$ typical of vigorously
star-forming local spirals.

The H$\beta$ equivalent widths measured here range from 8 \AA\ to over
200 \AA, with a median value of 44 \AA, typical of the late type spiral
and irregular galaxies surveyed by Kennicutt (1983).  Finally, in
Table~5 we list the production rate of ionizing photons, $Q_{Ly}$, for
each galaxy, assuming that all of the Lyman emission line photons get
scattered, re-processed, and re-emitted in the Balmer series (i.e. Case
B).  This Case B estimate of $Q_{Ly}$ should be regarded as a lower
limit since not all Lyman series photons will be re-processed,  and the
galaxy may be optically thin, in some directions, to Lyman continuum
photons (i.e. matter bounded rather than radiation bounded).

\section{Use of Chemical Analysis to Constrain Galaxy Evolution}

Many of the galaxies under consideration here have small half-light
radii, low dynamical masses inferred from the narrow emission line
widths, and high central surface brightnesses (Koo \etal\ 1995;  Guzman
\etal\ 1996), which make them characteristic of \HII\ galaxies (also
known as Blue Compact Galaxies; Searle \& Sargent 1972; Telles, Melnick
\& Terlevich 1997).  Some hypotheses predict that \HII\ galaxies evolve
into dwarf  or non-dwarf spheroidal galaxies after the starburst fades
(Bothun \etal\ 1986; Davies \& Phillips 1988; Telles, Melnick \&
Terlevich 1997).  However, 21-cm neutral hydrogen observations of
\HII\ galaxies show considerable rotational support 
(van Zee, Skillman, \& Salzer 1998) which is
inconsistent with the observed stellar kinematics of most dSphs (see
review by Mateo 1998).  Dwarf spheroidals also contain very little gas,
requiring that \HI--rich \HII\ galaxies  either consume or lose all of
their remaining gas if there is to be an evolutionary connection.

Knowledge of the chemical properties of star-forming galaxies at
intermediate redshifts constitutes an additional constraint on their
evolution.  Chemical data serve as useful constraints if the following
two simple principles are valid:  1) the heavy element content of
galaxies increases monotonically with age, and 2) the gas-phase
metallicity is at least roughly indicative of the metallicity of the
stellar population.  The first assumption should be valid except for
galaxies which accrete large quantities of gaseous or stellar material
with vastly different metallicity.  We can think of no reason why  this
first principle should not hold in the majority of galaxies.  Under
this assumption,  any evolutionary scenario linking \HII\ galaxies or
distant CNELGs with dSphs entails the boundary condition that the
metallicity of the final evolved product be equal to or greater than
its metallicity during the star-forming phase.

The second principle entails a variety of complex issues in galaxy
evolution, and it cannot yet be validated even for local galaxies.
However, we can be certain that large localized fluctuations in the
gas-phase metallicity in galaxies,  (perhaps cause by localized
chemical enrichment from starbursts, e.g.,  Kunth \& Sargent 1986;
Pagel, Terlevich, \& Melnick 1986) if they exist, would invalidate this
second principle.  Yet, a growing body of observational (Kobulnicky \&
Skillman 1997) and theoretical (Tenorio-Tagle 1996) evidence suggests
that heavy elements synthesized in bursts of star formation require
$>20$ Myr and possibly as long as a few Gyr to become mixed with the
surrounding interstellar gas.   Thus, the measured \HII\ region oxygen
abundances should be representative of the metallicities of the
recently-formed stars.  What about older stellar populations?  The
principle may not apply in galaxies which experience many episodes of
star formation, such as the Milky Way where stars span a range of over
3 decades in metallicity and more than 12 Gyr in age (review by
Wheeler, Sneden \& Truran, 1989).  For large galaxies with complex star
formation histories, there exists no single metallicity which
characterizes a whole galaxy.  Global nebular spectra are weighted
toward the brightest \HII\ regions, while global continuum (i.e.
light-weighted stellar) spectra are dominated by youngest stellar
population:  OB stars in the case of starbursts, A stars in the case of
postburst galaxies, and K giants in the case of old stellar populations
found in elliptical galaxies.  Despite the ambiguity of multiple
stellar populations, a weaker form of the second principle can still
serve as a guide in constraining the evolution of galaxies.  We posit,
2a) the gas-phase metallicity represents  an upper limit on the mean
stellar metallicity, and is indicative of the metallicity of the
youngest stellar population.

With these guiding principles in mind, we have plotted in Figures~4 and
5 diagnostic diagrams involving the chemical abundances, blue
luminosities, and velocity dispersions of program objects along with
local galaxies for comparison.  Figure~4 shows the oxygen abundances
versus absolute blue magnitude, $M_B$.  Objects from the CNELG sample
appear as filled circles with error bars.  The three sources with
direct oxygen abundance measurements appear a second time with a dotted
line connecting the empirical and direct O/H determinations which are
plotted as lower limits.  ELGs appear as open circles while the four
field objects appear as inverted filled triangles.  Local dwarf
irregulars as tabulated by Richer \& McCall (1995) appear as open
squares.  Local \HII\ galaxies from Telles \& Terlevich\footnote{We
adopt the \HII\ galaxy oxygen abundances from Telles \& Terlevich
(1997) and we convert their $M_V$ to $M_B$  assuming B-V=0.4 (Thuan
1983)} (1997) appear as stars.   Local spiral galaxies from ZKH are
plotted with asterisks using the mean oxygen abundance at 0.4 isophotal
radii.  Absolute magnitudes for the spirals are from Tully (1989)  but
scaled to $H_0$=50 \kms\ Mpc$^{-1}$ for consistency with the ELG
luminosities from Table~1.  Local spheroidal galaxies, NGC~205,
NGC~185, and Fornax, appear as open triangles at the oxygen abundance
determined from their planetary nebulae by Richer \& McCall (1995).  We
also include the Sagittarius dwarf with a planetary nebula abundance
measurement by Walsh \etal\ (1997).  Planetary nebulae provide the only
means of measuring the oxygen abundance in dSph objects.  However,
the derived O abundances reflect the metallicity of the planetary
nebula progenitors, which are likely to be the youngest, most metal
rich stars in the galaxy.   As pointed out by Walsh \etal\ (1997), the
planetary nebulae abundances represent an upper envelope on the
metallicity of the dominant stellar population.  In an effort to
characterize the metallicity of the stellar population,  we have also
estimated the mean {\it stellar} oxygen abundance for NGC~205 and
NGC~185 and a more luminous Sph, NGC~3605, using Mg (as a proxy for
oxygen) and Fe line indices reported by Trager \etal\ (1998), in
conjunction with the model stellar populations from Worthy (1994; see
Appendix).  Since Mg, like O, is an $\alpha$--process element and is
expected to be produced in constant proportion to oxygen, we use assume
that [Mg/H]$\simeq$[O/H].\footnote{We use the standard notation that
[Mg/H] = log(Mg/H) - log(Mg/H)$_{\odot}$.}  Although the observational
errors and uncertainties in the models are roughly 0.3 dex, we find
that the data for NGC~185 and NGC~205 are consistent with [Mg/H]=$-0.8$
and [Mg/H]=$-0.9$ respectively.  Thus, in Figure~4, we plot a second
point for NGC~185 at 12+log(O/H)=8.09 and NGC~205  at 12+log(O/H)=7.99
which is equivalent to [O/H]=$-0.8$ and [O/H]=$-0.9$ for the standard
abundance distribution (Anders \& Grevesse 1989).  The spheroidal
NGC~3605 (Trager \etal\ 1998) is plotted using the same procedure to
convert from [Mg/H] to [O/H].

\subsection{The Evolution of the Mass-Metallicity Relationship
with Cosmic Epoch}

Figure 4 shows the luminosity-metallicity relation obeyed by nearly all
local galaxies (e.g.,  Lequeux \etal\ 1979; French 1980; Faber 1973;
Brodie \& Huchra 1991; ZKH; Richer \& McCall 1995).  The solid line
shows a least squares fit to the local irregular and spiral galaxies.
The figure illustrates that, as pointed out by Skillman (1992), only
\HII\ galaxies deviate systematically from the luminosity-metallicity
trend.  They appear 0--3 magnitudes brighter than objects of similar
metallicity, as might be expected if a strong starburst has temporarily
lowered their mass/light ratios.  Furthermore, strong nebular emission
lines may artificially increase the broadband luminosities
of \HII\ galaxies, especially in V-band by as much as 1 mag.

Figure~4 reveals that all objects observed in this program fall along
the main metallicity-luminosity correlation.  The least luminous
object, L2-410083, is also the most metal poor.  It lies near points
representing large irregular galaxies like NGC~1569 or the LMC.  The
most luminous objects in the sample are the most metal-rich.  They
occupy a similar locus in the metallicity-luminosity correlation as
local spiral galaxies, although spirals show a dispersion of $\pm$0.3
dex at a given luminosity and are known to have significant internal
chemical variations.  Formally, the slope of the mass metallicity
relation for the distant objects is $-$0.128$\pm$0.020 dex mag$^{-1}$,
consistent with  $-$0.144$\pm$0.003 dex mag$^{-1}$ for the local
galaxies drawn in Figure~4.  The zero point offset between the two
samples is consistent within the $1\sigma$ errors as well.  A rigorous
measurement of the luminosity-metallicity relation at earlier epochs
will require observations of intrinsically underluminous (dwarf;
$M_B<-18$) objects at redshifts comparable to those of the brighter
objects in the sample ($z\sim0.4$).  Due to magnitude-limited selection
effects in most redshift surveys, very few dwarf galaxies are known at
$z>$0.2.    Furthermore, we specifically selected the objects for this
survey on the basis of their strong emission lines, and the sample
comprises a range of lookback times.  Thus, it would be inappropriate
to draw strong conclusions regarding the cosmic chemical evolution of
normal galaxies from this small, statistically biased sample.

With the above caveats in mind, we compare the luminosity-metallicity
relation of the 13 objects with  3.2---6.4 Gyr look back times 
to the relation followed by local galaxies (i.e., spirals) 
with comparable data.  In
order to construct a suitable local comparison sample, we compute
the nebular oxygen abundances for 27 local spiral galaxies using their
global (i.e., spatially integrated) emission line spectra.  We begin
with raw global galaxy spectra from Kennicutt (1992---K92; 14 galaxies)
and pseudo-global spectra constructed by summing the spectra of
\HII\ regions observed by ZKH in 13 galaxies for which the data were
available in electronic form.  We compute empirical oxygen abundances
using strong line ratios as described in Section~2 for each galaxy.
Absolute blue magnitudes are taken from K92 and ZKH but scaled from
$H_0$=75 \kms\ Mpc$^{-1}$ assumed in those works to  $H_0$=50
\kms\ Mpc$^{-1}$.  Figure~5 shows an expanded version of the
luminosity-metallicity relation for spirals, using the oxygen
abundances derived from global spectra.  The dashed line shows a least
squares fit to the spirals, 12+log(O/H)=$-$0.108$\pm$0.040 $M_B$ +
6.637$\pm$0.841.  Solid symbols represent the CNELGs and open circles
represent the ELGs, as in the previous figures.  The entire galaxy
sample is roughly consistent with the local relation, given the intrinsic
scatter of $\sim$0.3 dex in log(O/H) among local spirals at a given
magnitude.  A formal linear fit to the 9 CNELG/ELG points is consistent
with the best fit line for spiral galaxies, but there is a tendency for
the distant sample to lie below that dashed line.  There is marginal
evidence for the CNELGs and ELGs to be 0.1--0.2 dex more metal poor
than local spirals at a given luminosity, or, alternatively, for them
to be 1--3 magnitudes brighter than the mean spiral with comparable
metallicity.  Figure~6 shows a histogram of the oxygen abundance
residuals of the 13 galaxies compared to the linear fit to the local
spirals.  The histogram for the local spirals centers around zero,
while the distant galaxies show a mean near $12+log(O/H)=-0.1$.  Given
the 0.3 dex spread in O/H at a given luminosity, this constitutes only
marginal evidence (1.1$\sigma$ in the error of the mean) for evolution
of the luminosity-metallicity relation, since the targets observed here
are a highly-selected sample of objects.  However, the close
correspondence of the distant sample at lookback times of 3.2---6.4 Gyr
to the local objects does indicate that no more than 1--3 magnitudes of
fading, and no more than 0.2--0.3 dex of oxygen enrichment can occur
without causing noticeable deviations from the luminosity--metallicity
relation. Overall, the 9 objects at lookback times of $\sim$5.5 Gyr, do
not probe far enough into cosmic history to expect significant
signatures of chemical evolution.  Chemical measurements of more
distant star-forming galaxies should provide such evidence.
If it can be shown that the same luminosity-metallicity
and mass-metallicity relations apply at
very early epochs, then  
oxygen abundance measurements may serve as surrogates for
luminosity and mass indicators when the latter are difficult to
measure directly.

\subsection{The Descendants of the Emission Line Galaxies}

The physical characteristics of the descendents of the ELGs and CNELGs
will depend on their subsequent luminosity and chemical evolution.
Barring a major merger or accretion event, their gas-phase oxygen
abundances can only increase as star formation
proceeds and gas is turned into stars.  Their luminosities will presumably
decrease, but the amount of fading depends on the underlying stellar
population.  Guzman \etal\ (1998) have estimated that if the current
burst involves 10\% of the mass of the stellar component, then the
CNELGs will fade by 2-4 magnitudes in $\sim$5 Gyr.  The amount by which
the {\it gas-phase} metallicity can increase during subsequent evolution
depends upon the present gas mass fraction.  The amount by which the
mean {\it stellar} metallicity will increase also depends upon the
present gas mass fraction, and the present stellar mass fraction.
Qualitatively, if most of the baryonic mass is already locked up in
stars, then the final systemic metallicities will not increase
significantly from their current values.  The present bursts could be
the last major episodes of star formation if all of the gas is turned
into stars, or if the remaining gas is blown away in galactic winds.
Based on the large ionized gas masses in these systems (several
$\times10^{6}$ \mo) we infer the presence of many times this mass in
atomic and molecular gas based on results obtained in local
star-forming galaxies.  Thus, there is clearly sufficient raw 
material for subsequent star formation.  The present star formation
episode is unlikely to be the last unless the remaining gas is expelled
in galactic winds.  However, the possibility that vigorous starbursts
expel a large fraction of the atomic gas in galactic winds (Mathews \&
Baker 1971; Larson 1974; Dekel \& Silk 1986) now looks increasingly
improbable (Marlowe \etal\ 1995; Skillman 1997; Martin 1998; MacLow \&
Ferrara 1998) except perhaps in the smallest dwarf galaxies ($M<10^{7}$
\mo).

Regardless of uncertainties on the subsequent luminosity and chemical
evolution, it is clear from Figure~4 that 7 of the 13 objects in our
sample, including two of the CNELGs, have gas phase oxygen abundances
higher than the planetary nebula oxygen abundances observed in dwarf
spheroidals like NGC~205.  Since the PNe abundances represent an upper
envelope to the metallicity attained by the most metal-rich population
in NGC~205, these 7 objects are already too chemically enriched to
become an NGC~205.  They must evolve into higher-metallicity, and in
keeping with the luminosity-metallicity correlation, higher-luminosity
systems.  If the two most metal-rich CNELGs, L2-408115 and SA68-206134,
fade enough to become spheroidal galaxies, they must become more
luminous, metal-rich objects such as NGC~3605.

The other six objects in our sample exhibit oxygen abundances equal to,
or below the planetary nebulae oxygen abundances in NGC~205.  From a
chemical standpoint, the two most metal-poor CNELGs in our sample,
L2-410083 and L2-411500, could evolve into objects like NGC~205 provided
that they fade by $\sim$1 and 4 magnitudes, respectively, {\it and}
provided that their gas phase oxygen abundances do not increase by more
than 0.35 and 0.15 dex, respectively, during subsequent evolution.

It is evident from Figure~4 that the luminosity-metallicity correlation
based on planetary nebulae in four spheroidal galaxies (open triangles)
is offset from the mean \HII\ region relation by $\Delta{O/H}\simeq$0.5
dex.  This deviation is consistent with a chemical evolution scenario
in which the bulk of the stellar population formed from gas at a lower
metallicity (measured by the integrated Mg line indicies) compared to
final generation of stars which become the progenitors of
presently-observed planetary nebula.  The youngest, most recent
generation of stars now forming planetary nebulae would have been born
at a time when the (nearly exhausted?) gas supply had become more
chemically enriched.  Thus, it is expected that the gas-phase
metallicity, or the metallicity of the youngest stars, should be
greater than the mean stellar metallicity.

\subsection{Dynamical-Chemical Diagnostics}

Figure~7 shows the oxygen abundance versus the velocity dispersion,
$\sigma_v$, for a subset of the galaxies from Figure~4 which have
measured kinematics.  For local irregular and spiral objects from
Richer \& McCall (1995) and ZKH, we compute $\sigma_v$ from the
single-dish \HI\ line width at 20\% peak
intensity, $W_{20}$, compiled in Huchtmeier \& Richter (1989)
using the prescription of Tully \& Fouqu\'e (1985),

\begin{equation}
W_R^2 = W_{20}^2 + W_t^2 - 2W_{20}W_t[1-e^{-(W_{20}/W_c)^2}] - 2W_t^2
e^{-(W_{20}/W_c)^2}.
\end{equation}

\noindent Here, $W_t=38$ \kms\ is the width due to turbulent motions
and $W_c=120$ \kms\ is the transition point between galaxies having
Gaussian versus double-horned \HI\ profiles, as in Tully \& Fouque
(1985).  We compute $\sigma_v$ from the the rotation full-amplitude,
$W_R$, using $\sigma_v=0.5W_R/1.4$.  When more than one
\HI\ measurement is listed in Huchtmeier \& Richter, we adopt the
median, and use the dispersion among the various measurements as an
estimate of the 1$\sigma$ uncertainty.  Velocity dispersions for the
ELGs are taken directly from Guzman \etal\ (1996) and Koo
\etal\ (1997), while data for the dwarf spheroidal systems come from
Bender, Paquet \& Nieto (1991), Peterson \& Caldwell (1993), and
Bender, Burstein, \& Faber (1993).  The data in Figure~7 exhibit a
strong correlation between velocity dispersion and oxygen metallicity.
Since both optical luminosity and velocity dispersion should trace the
underlying stellar and dynamical masses respectively, the strong
correlations in both Figures~4 and 7 are consistent with a
mass-metallicity relationship for galaxies of all types.  It is
interesting to note that \HII\ galaxies, which were discrepant in the
luminosity--metallicity diagram, are indistinguishable from
irregular galaxies on this velocity dispersion--metallicity diagram.
This suggests the hypothesis
that \HII\ galaxies are indeed overluminous for their
metallicity due to a strong starburst (and broadband contamination by
emission lines), rather than being under-enriched for their luminosity
and mass.

The two most luminous ELGs, SDG~125 and SDG~146, which also have the
largest velocity dispersions, fall among the local spiral galaxies in
Figure~7.  The remaining 7 objects cluster along
the boundary between the spiral and dIrr populations.  As a group, they
lie slightly above the O/H---$\sigma_v$ relation for galaxies of
similar velocity dispersion.  However, several observational effects
could cause the measured H$\alpha$ line width to be systematically
low.  Since inclinations are not known for the sample, their positions
along the $\sigma_v$ axis are actually firm lower limits; correction by
a factor of $1/sin(i)$ would move them to the right in Figure~7. A
(statistically) typical correction factor of $1/sin(60\deg)\simeq$ would
shift all of these points to the right by 15\%.  Furthermore, velocity
dispersions derived from optical emission lines arising in the warm
ionized gas are likely to underestimate $\sigma_v$ derived from the
more spatially--extended neutral atomic (\HI) gas.  Rix \etal\ (1997)
perform Monte-Carlo simulations to characterize the effect of viewing
angle, seeing, dust extinction, and clumpy gas distributions on the
measured optical emission line profiles.  They find a mean correction
factor of $\sim$1.4 which would bring the observed galaxies into
consonance with the metallicities and velocity dispersions of local
spirals and large irregulars.  Lehnert \& Heckman (1996) find that in
starburst galaxies, the emission-line region traces only a fraction of
the solid body portion of the rotation curve.  We conclude that the
dynamical-chemical evidence is consistent with the proposition
that ELGs and CNELGs follow the same velocity-dispersion--oxygen abundance
relation as local star-forming spiral and irregular galaxies.

The small half-light radii of the three CNELGs, however, seem to rule
out the existence of an underlying disk in these objects (Guzman
\etal\ 1996, 1998).  Even though the three CNELGs are chemically
similar to the ELGs, their structural parameters suggest a
fundamentally different evolutionary path.  We reproduce in the left
panel of Figure~8 the relations between half-light radius, $R_e$, and
velocity dispersion, $\sigma_v$, shown by Guzman (1996).  Dotted
polygons outline the locus occupied by local spiral, irregular,
elliptical and spheroidal galaxies.  The ELGs fall among the spiral and
irregular populations, while two of the three CNELGs lie among the
spheroidal galaxies.  In the right panel of Figure~8 we show the
relation between surface brightness inside the half-light radius versus
the absolute magnitude of the ELG/CNELG sample.  Dashed lines denote
the regions occupied by local galaxy types.  An arrow indicates the
direction of passive evolution due to fading of the starburst
population.   Except for SDG~223, the ELGs predominantly occupy the
region of local spiral and irregular galaxies.  The CNELGs, and
SDG~223, occupy a region consistent with an elliptical or (with fading)
spheroidal classification.

\section{Summary: What are Distant ELGs?}

The distant galaxies in this study span a range of velocity dispersions
($\sigma_v=$29 \kms\ --- 200 \kms), magnitudes ($M_B=-$16.6 --- 22.0)
and blue half-light radii (1.22 kpc --- 6.7 kpc).  Clearly, they
represent a diverse group of objects which, in all likelihood, had
different origins and will have different evolutionary descendants.
Nevertheless, the targets share many common features.  Their
metallicities, luminosities, and velocity dispersions (with small
statistical upward adjustments for unknown inclination) are consistent
with those of local spiral or the most massive irregular galaxies.
They show vigorous star formation similar to the most actively bursting
spiral and blue compact galaxies.  They contain several $\times10^{6}$
\mo\ of ionized gas, suggesting that the mass of atomic and molecular
gas may exceed this by several orders of magnitude based on
ionized/neutral gas ratios in local galaxies.  All of the observed
objects fall along the luminosity-metallicity relation established by
local galaxies.   Only \HII\ galaxies and spheroidal galaxies deviate
from the local luminosity-metallicity trend.  Hence, these CNELGs do
not appear to be distant counterparts of \HII\ galaxies, even though
they fall along the same H$\beta$ versus velocity dispersion relation
as local \HII\ galaxies (see Figure~3 of Koo \etal\ 1995).  Based on
their N/O ratios, they have experienced considerable secondary nitrogen
production similar to local spiral galaxies.  Since secondary N
production is believed to occur only in intermediate mass (4--8 \mo)
stars (Renzini \& Voli 1981) these galaxies must have experienced at
least one, and probably several prior episodes of star formation in
order to display secondary nucleosynthsis products.  This is consistent
with the presence of older underlying stellar populations and extended
exponential (i.e., disk-like or spheroidal-like) surface brightness
profiles (Guzman \etal\ 1998).  Such an underlying stellar population
limits the amount of fading which can occur after the end of the
starburst episode.

{\it The CNELGs:} The four CNELGs in this study differ from the rest of
the sample most significantly in their small half-light radii and high
central surface brightnesses as shown in Figure~8.  These
characteristics make them most similar to local spheroidal or
elliptical galaxies.  CNELGs may evolve into spheroidal galaxies once
the starburst subsides provided that they lose or consume their
remaining gas.  Based on the tightness of the luminosity-metallicity
correlation for star-forming galaxies, we predict they will not fade by
more than 1-2 magnitudes unless they violate the local
luminosity-metallicity relation, as well-studied spheroidals appear to
do, based on their planetary nebula oxygen abundances.  Two of the
CNELGs, L2-410083, and L2-411500 are sufficiently metal poor that fading
by 1 and 4 magnitudes, respectively, could make them consistent with
the luminosity and planetary nebula oxygen abundances of NGC~205.  The
two other CNELGS, L2-408115 and SA68-206134, are presently more oxygen rich
than the planetary nebulae in NGC~205.  They cannot evolve into such an
object unless, contrary to chemical evolution principles, they become more
metal-poor as they age.  If these CNELGs evolve into spheroidal
objects, then we suggest more luminous and chemically-rich spheroidals
like NGC~3605 as local examples of their descendants.  An alternative
hypothesis, equally consistent with the chemical data, is that the
CNELGs are the precursors to spiral bulges seen during an episode of
vigorous star formation before their disks have formed.  Their small
emission line widths may be an artifact of the starburst region
sampling only the inner part of the galaxy's potential well (Lehnert \&
Heckman 1996).

{\it The ELGs:}  The six ELGs in this study are chemically and
morphologically consistent with spiral galaxies during an episode of
intense nuclear star formation.  They have
larger half-light radii than the CNELGs, ranging from
2 to 6 kpc, consistent with substantial underlying stellar 
disks.  SDG~223 is an intermediate object, with
a central surface brightness and a half light radius between the
compact CNELGs and the other ELGs which show more extended disk-like
morphologies.  SDG~223 also has the highest central surface
brightness of the ELGs, making it more typical of 
elliptical galaxies.  

{\it Other Field Galaxies:} The four objects we call ``field galaxies''
have small emission line equivalent widths compared to the ELGs and
CNELGs.  They have correspondingly lower star formation rates of 1--5
\mo\ yr$^{-1}$based on their H$\beta$ luminosities.  However, they are
equally, or perhaps more highly chemically enriched.  Unfortunately,
their morphological and dynamical properties are not yet measured.
Their spectra closely match those of the Sb and Sc galaxies in
Kennicutt (1992), leading us to tentatively classify them as spiral
galaxies much like those observed in the local universe.

At the moment, this chemical study contains only 3 CNELGs in common
with the Guzman \etal\ (1998) sample which have measured velocity
dispersions and morphologies.  Firm conclusions about the evolution of
the ELGs and CNELGs at intermediate redshifts will require complete
chemical, dynamical, and morphological parameters for a larger sample
of objects.  Intrinsically under-luminous galaxies at those redshifts
are especially interesting.  At the moment, however, surveys at
redshifts in excess of $z=0.2$ are not sensitive to dwarf galaxies
($M_B>-18$) so chemical and dynamical investigations are limited to the
bright end of the luminosity function.   Comparison of
intermediate--$z$ dynamical properties to objects in the nearby
universe would benefit from a large sample of velocity-resolved, {\it
spatially-integrated} emission-line spectra of local galaxies.  Linking
the intermediate--$z$ star-forming objects with the evolutionary
descendants will also require a better understanding of the differences
between $\alpha$-process and Fe-peak elements in the gaseous and
stellar constituents of the Local Group.

\acknowledgments   We are grateful to Drew Phillips for his expertise
in preparing the technical details for the observing run, and Luc
Simard for help with EXPECTOR during reductions.  Rafael Guzman
provided coordinates for L2-410083 which enabled us to extend this
study to a system with lower luminosity and metallicity.  We thank
Sandy Faber, Don Garnett, Rafael Guzman, David Koo, Evan Skillman,
Elena Terlevich, and Guy Worthey for inspirational and educational
conversations.  D.~Z.  gratefully acknowledges financial support from a
NASA LTSA grant (NAG-5-3501) and the David and Lucile Packard
Foundation.  Data presented herein were obtained at the W.~M.  Keck
Observatory, which is operated as a scientific partnership among the
California Institute of Technology, the University of California and
the National Aeronautics and Space Administration.  The Observatory was
made possible by the generous financial support of the W.M. Keck
Foundation.  H.~A.~K appreciates hospitality at the Aspen Center for
Physics during the June 1998 summer workshop on star formation and at
the 1998 Guillarmo Haro International Program for Advanced Studies in
Astrophysics at the Instituto National de Astrofisica Optica y
Electronica (INAOE) in Puebla, Mexico where this work was completed.
H.~A.~K receives support from NASA through grant \#HF-01094.01-97A
awarded by the Space Telescope Science Institute which is operated by
the Association of Universities for Research in Astronomy, Inc. for
NASA under contract NAS 5-26555.

\appendix
\section{Use of Stellar $Mg~b$ Indices to Estimate $\alpha$-element Abundances
in Dwarf Spheroidal Galaxies}

There exist very few galaxies of any type for which the  chemical
properties of both the stellar and gaseous components are known.  The
most reliable stellar abundance data comes from Fe absorption lines
which are easily observed in stellar atmospheres.  In the gas phase,
collisionally-excited emission lines of singly and doubly ionized
oxygen in \HII\ regions provide the most reliable metallicity
estimates.  If Fe and O both had a common nucleosynthetic origin,
measurement of either element would suffice as a fiducial indicator of
the overall system metallicity.  However, in our own Galaxy, large O/Fe
variations exist between globular clusters, the Galactic bulge, and the
disk, presumably due to the varying chemical contributions of Type II
($\alpha$-process elements) versus Type Ia (Fe-peak elements)
supernovae (e.g., review by Wheeler, Sneden, \& Truran 1989).

A fair comparison of the oxygen abundance measurements of the
intermediate-redshift star-forming objects observed in this work to the
$\alpha$-element abundances in dwarf spheroidals requires a measurement
of the stellar $\alpha$-element content in these local galaxies.
Richer \& McCall (1995) obtained the first such data,
 measuring the oxygen abundances of planetary nebulae in NGC 205,
NGC~185, and Fornax.  Comparison with stellar iron abundances derived
from color-magnitude diagrams and theoretical isochrones led them to
conclude that the [O/Fe] ratio in dwarf spheroidals is systematically
0.3-0.4 dex above the [O/Fe] ratio in dwarf irregular galaxies like the
Magellanic Clouds.  This elevated $[\alpha/Fe]$ ratio is consistent
with a short star formation timescale for the bulk of the stellar
population, and it suggests that dwarf spheroidals   experience
fundamentally different star formation histories than dwarf irregular
galaxies.  Thus, they argue, an evolutionary link between the two is
unlikely.  However, the oxygen abundance measurements come from
planetary nebulae progenitors which are comparatively young and
massive.  They represent the high-metallicity tail of the stellar
abundance distribution, and so the derived oxygen abundances are best
regarded as an upper limit.

As another means of estimating the $\alpha$ element metallicity in
dwarf spheroidal galaxies, we consider the spectroscopic metal line
indices of NGC~205, NGC~185, and NGC~3605 reported in Trager
\etal\ 1998.  In Figure~9 we plot the H$\beta$ index versus $Mg b$ and
$Fe5270$ indices for these objects, using different symbol sizes to
denote the small and large aperture observations.  H$\beta$ is
primarily an age indicator, while the Fe and Mg indices are primarily
metallicity indicators.  The abundance of Mg is of special interest
since, as an $\alpha$-process element, it shares a common
nucleosynthetic origin with O.  Mg traces the O abundance in all known
environments except possibly in the Galactic bulge (McWilliam \&
Rich 1994).  Figure~9 also includes model grids of age and metallicity
from Worthy (1994) for single-burst stellar populations with ages of
17, 12, and 8 Gyr and metallicities corresponding to
$[Fe/H]=-2.0$ --- $-0.25$.  In the left panel we show the data and
grids for the $Fe5270$ index.  The data points for NGC~205 and NGC~185
generally lie in a region consistent with metallicities of
$[Fe/H]=-1.0$ --- $-0.5$, but they span wide range in age.  NGC~3605
and two measurements of NGC~205 through the smallest apertures
(corresponding to the innermost regions of the galaxy) lie off the
model grids, suggesting very young ages, but unfortunately models do
not yet yet exist for metal-poor stellar populations younger than 8
Gyr.  In the right panel we show the data and grids for the $Mg b$
index.  In this panel, NGC~185, and the two largest aperture
measurements of NGC~205 lie between the $[Fe/H]=-1.0$ and $[Fe/H]=-1.5$
models.   Based on the Mg data, these dwarf spheroidal would appear to
be more metal deficient than the Fe data would indicate.  However,
caution must be exercised in the interpretation because the models are
calibrated empirically from globular cluster line indices rather than
stellar models using the standard solar abundance distribution.   Model
lines of constant metallicity cannot be used to directly infer the
ratio of ratio of $[Mg/H]$, or by proxy, $[\alpha/H]$, without taking
into account systematic chemical variations in the objects used to
calibrate the models.  Metal-poor globular clusters with
$[Fe/H]\le-0.5$ show a nearly constant overabundance of
$\alpha$-process elements compared to the solar value,
$[\alpha/Fe]\sim0.4$ (see review by Carney 1996).  Based on this mean
$[\alpha/Fe]$ overabundance in globular clusters on which the models
are calibrated, we apply a correction of +0.4 dex to the model grids in
order to estimate the true $[Mg/H]$ values for the dwarf spheroidals.
A smaller font shows the estimated $[Mg/H]$ values for each model in
the right panel of Figure~9.  The data points for NGC~185 and the two
large-aperture observations for NGC~205 lie in the range
$-0.9<[Mg/H]<-0.6$.

Based on the model grids in Figure~9, we adopt mean values for NGC~185
and NGC~205 (two largest apertures only) of [Fe/H]$=-1.1$ and
[Fe/H]$=-0.8$ respectively.  For [Mg/H] we adopt $-0.8$ and $-0.9$
respectively.  The corresponding [Mg/Fe], and by proxy, [$\alpha$/Fe]
values are +0.3 and -0.1 for NGC~185 and NGC~205 with probable errors
of $\pm$0.2 dex.  Assuming that O and Mg are both representative of
$\alpha$-process elements, these [Mg/H] ratios are fully consistent
with the [O/Fe] estimates of $+0.13\pm0.19$ and $-0.05\pm0.32$ inferred
for the two galaxies by Richer \& McCall (1995) after correcting the
PNe oxygen abundances for chemical evolution between the time when the
Pne progenitors formed, and the epoch when the bulk of the stellar
population was formed.  Because the [Mg/H] measurements are derived
from the same stellar population as the [Fe/H] measurements and do not
require corrections for (largely unknown) effects of galaxy chemical
evolution, They may represent a more robust measure of the [$\alpha$/H]
ratio in dwarf spheroidals.  Nevertheless, the reasonable agreement
between the [$\alpha$/H] ratios derived by Richer \& McCall (1995) and
those found here provides confidence in the results of both
approaches.

For the purposes of this paper, and the discussion of \S~3.1, we
will use the [Mg/H] estimates of $-0.8$ and $-0.9$ to infer that the
mean stellar oxygen abundance is [O/H]=$-0.8$ and [O/H]=$-0.9$ in NGC~185 and
NGC~205 respectively.

\clearpage

\begin{figure}
\centerline{\psfig{file=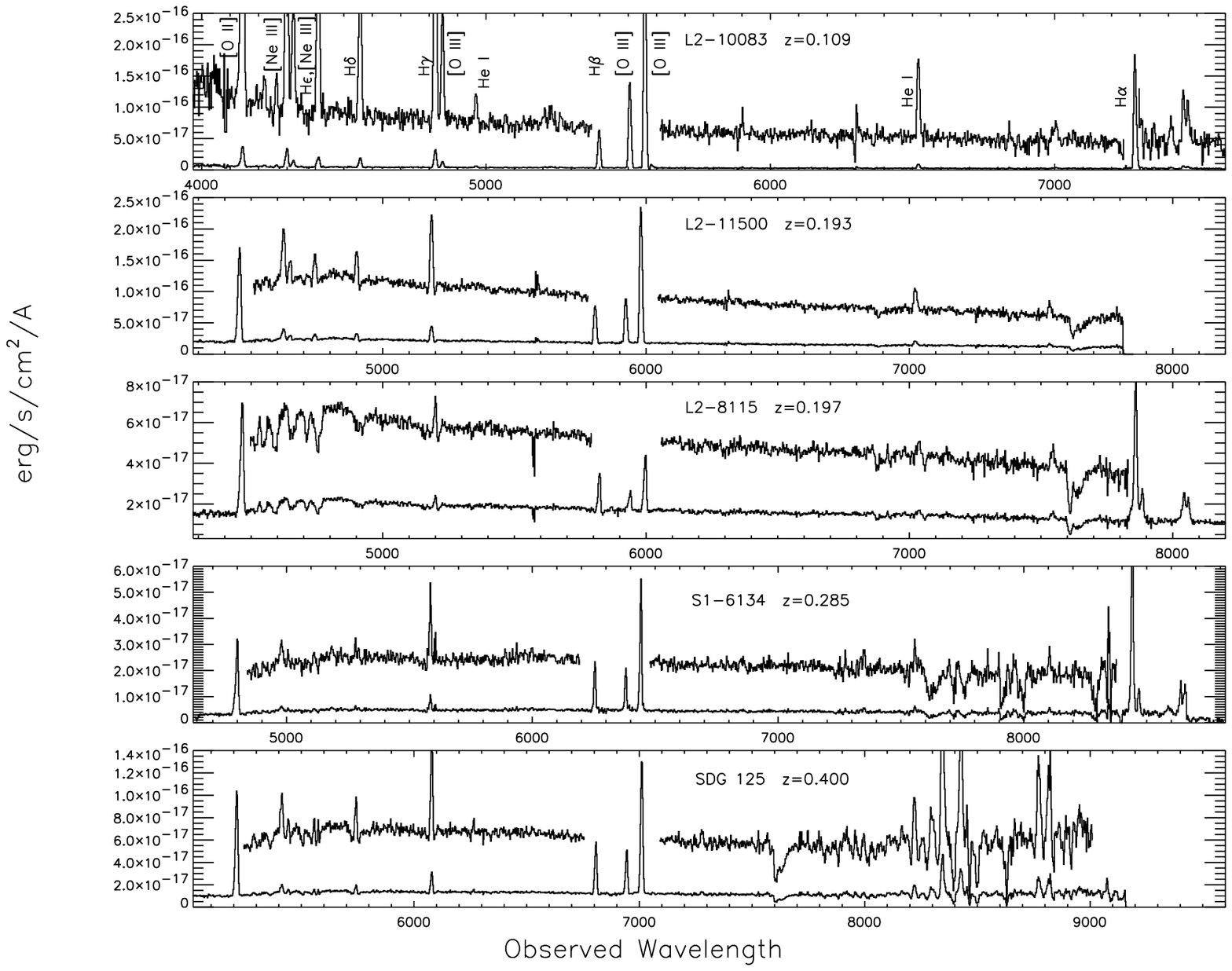,width=5.5in,angle=90}}
\end{figure}
\clearpage

\begin{figure}
\centerline{\psfig{file=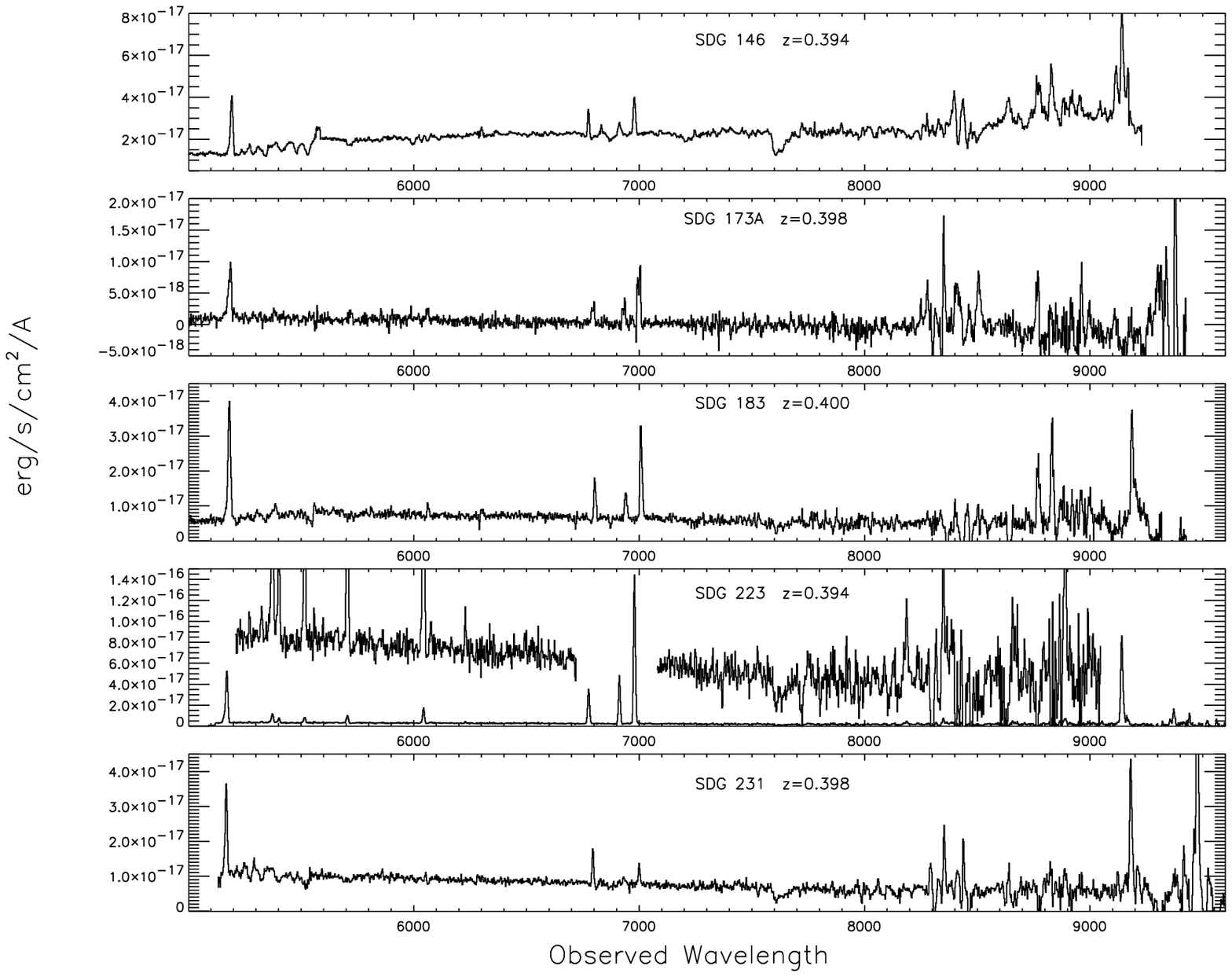,width=5.5in,angle=90}}
\end{figure}
\clearpage

\begin{figure}
\centerline{\psfig{file=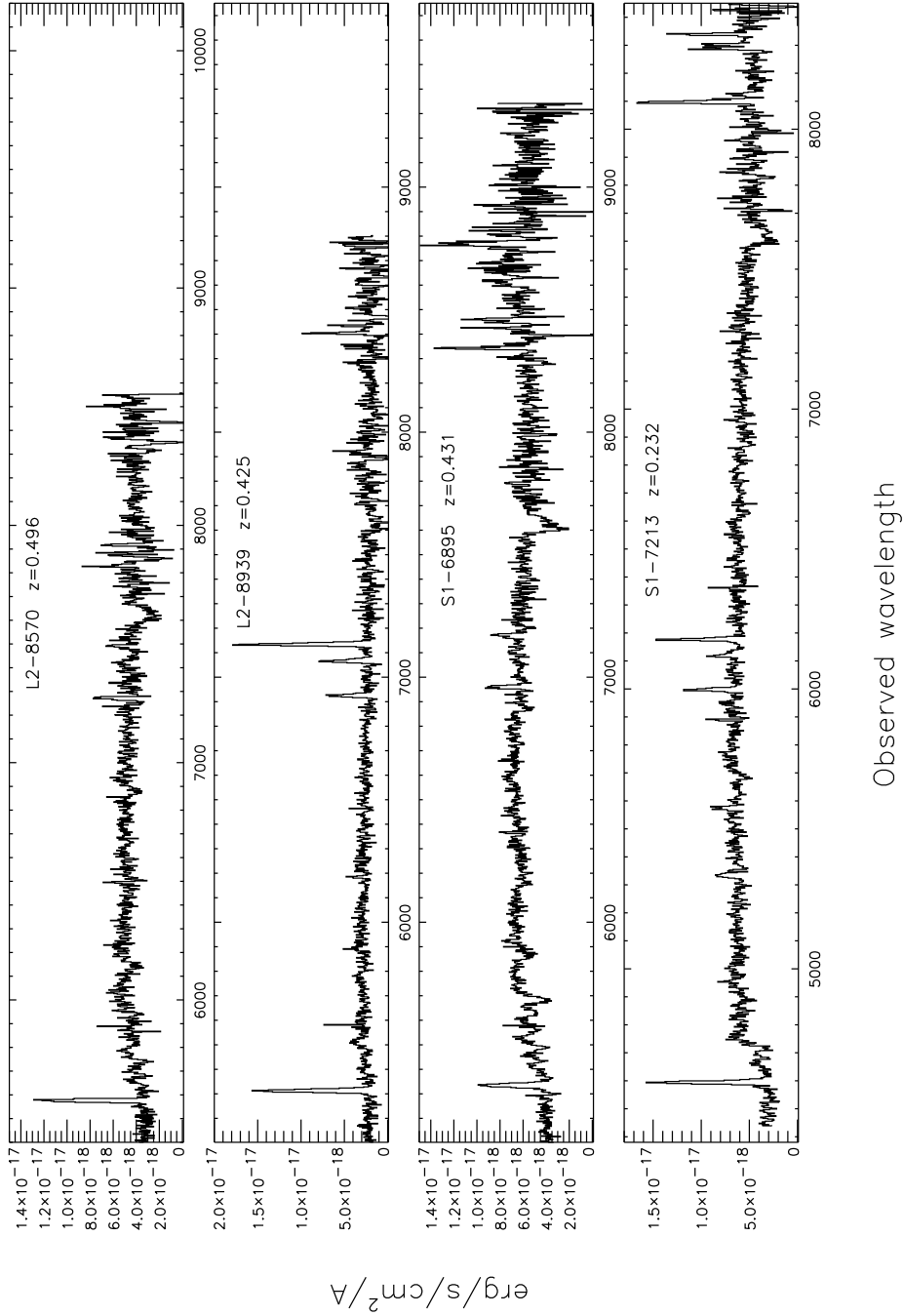,width=5.5in,angle=90}}
\figcaption[plotspec1.ps] {a,b,c: Keck LRIS spectra of emission line
galaxies and several field objects at $z=0.1$ to $z=0.5$.  The spectra
are not corrected for reddening or redshift.  Where the dynamic range
is large, the spectrum is scaled by an arbitrary factor and plotted a
second time for display purposes.  Labels on prominent emission lines
appear in the first panel.  \label{plotspec1} }
\end{figure}

\begin{figure}
\centerline{\psfig{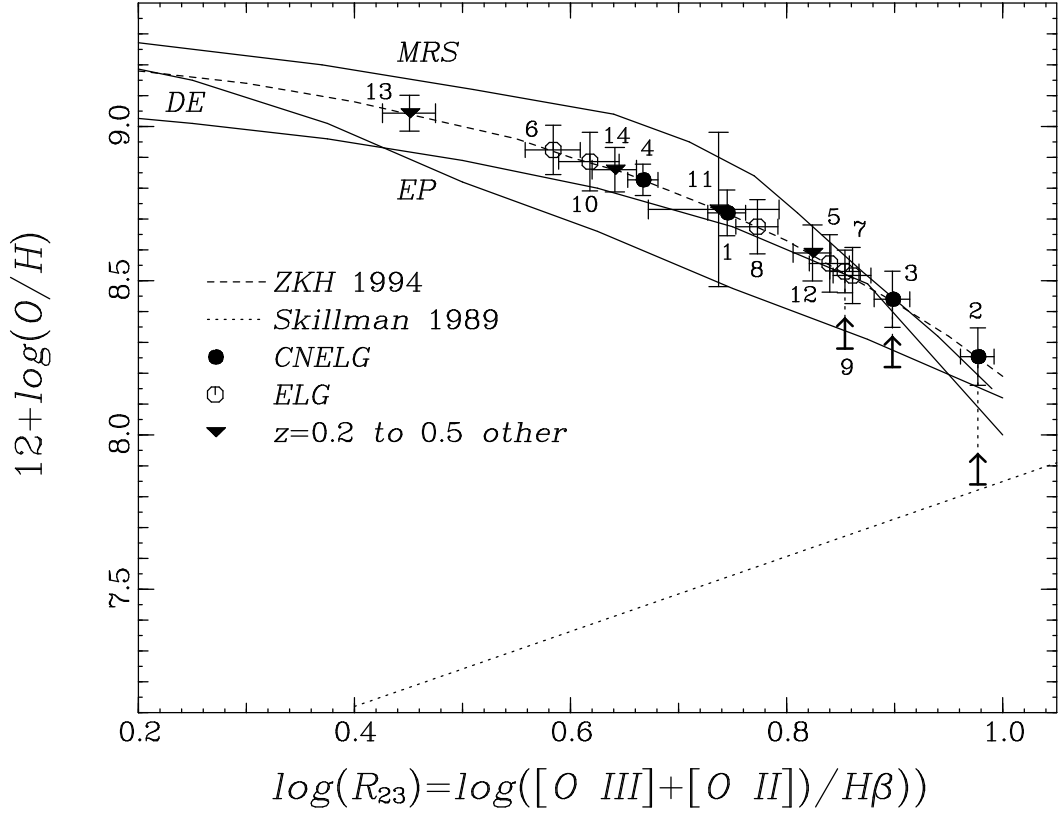}}
\figcaption[R23_OH.ps]{Calibrations of the strong-line ratio,
R$_{23}$=$(I_{3727} +I_{4959}+I_{5007})/H\beta$, to oxygen abundance,
12+log(O/H), in \HII\ regions.  Solid lines display the calibrations of
Edmunds \& Pagel (EP, 1984), McCall, Rybski \& Shields (MRS, 1985), and
Dopita \& Evans (DE, 1986).  The dashed line is the empirical
polynomial approximation to the aforementioned three studies by
Zaritsky, Kennicutt, \& Huchra (1994).  The dotted line shows the
low-metallicity branch of the $R_{23}$ calibration fitted by Skillman
(1989).  Symbols indicate the positions of CNELGs ({\it filled
circles}), other ELGs  ({\it open circles}), and four field galaxies
with emission lines ({\it filled triangles}).  Dotted lines connect
oxygen abundance measurements derived from the strong-line method with
a lower limit based on direct measurements of the electron temperatures
in three objects.  O/H derived from direct measurements should be
regarded as firm lower limits due to systematic effects discussed in
the text.   Numbers correspond to galaxy ID\#s in Table~1 for easy
identification.  \label{R23_OH} }
\end{figure}

\begin{figure}
\centerline{\psfig{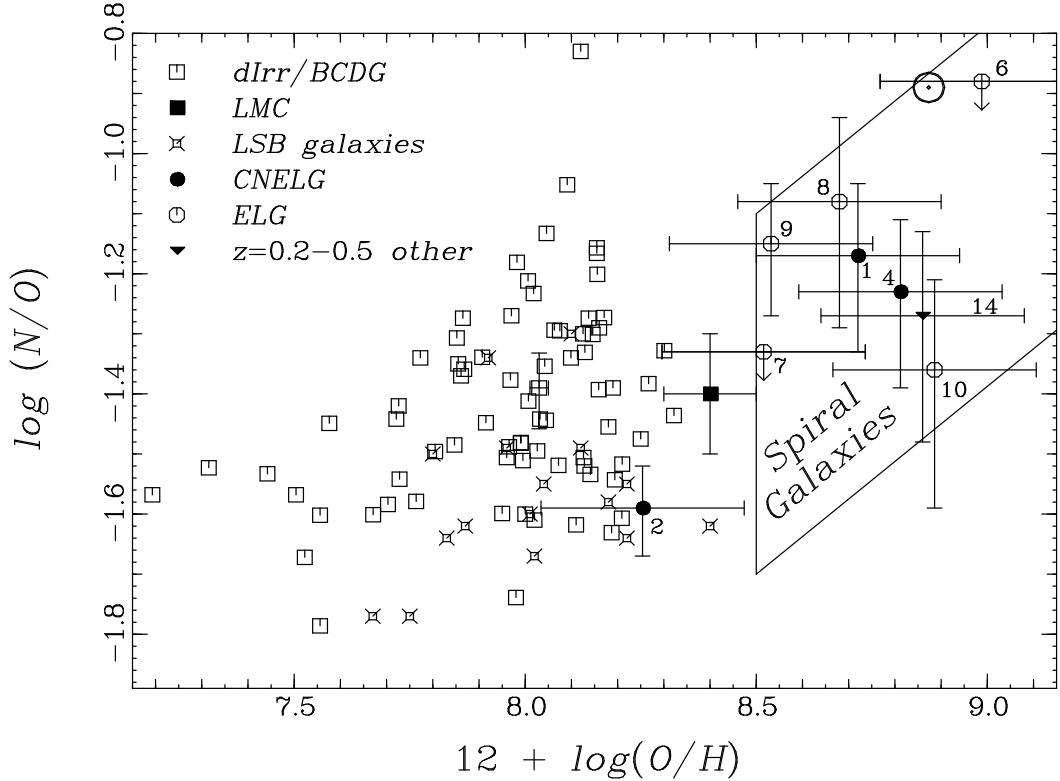}}
\figcaption[OH_NO.ps]{12+log(O/H) versus log(N/O) for the target
objects and an assortment local galaxies.   Symbols indicate the
chemical abundances of CNELGs ({\it filled circles}), ELGs  ({\it open
circle}), and field galaxies with emission lines ({\it filled
triangles}).  For comparison, we also show nearby dwarf irregular and
blue compact galaxies (open squares; from the compilation of Kobulnicky
\& Skillman 1996), the LMC (filled squares; error bars represent the
dispersion among LMC \HII\ regions; Pagel \etal\ 1978), and
low-surface-brightness galaxies (stars; van Zee 1998).  Numbers
correspond to galaxy ID\#s in Table~1 for easy identification.  With
the exception of L2-10083, all of the observed objects at $z>0.2$ show
relatively high oxygen abundances and evidence for secondary N
production $(log(N/O)>-1.3)$ which makes them chemically most similar
to local spiral galaxies.  \label{OH_NO} }
\end{figure}

\begin{figure}
\centerline{\psfig{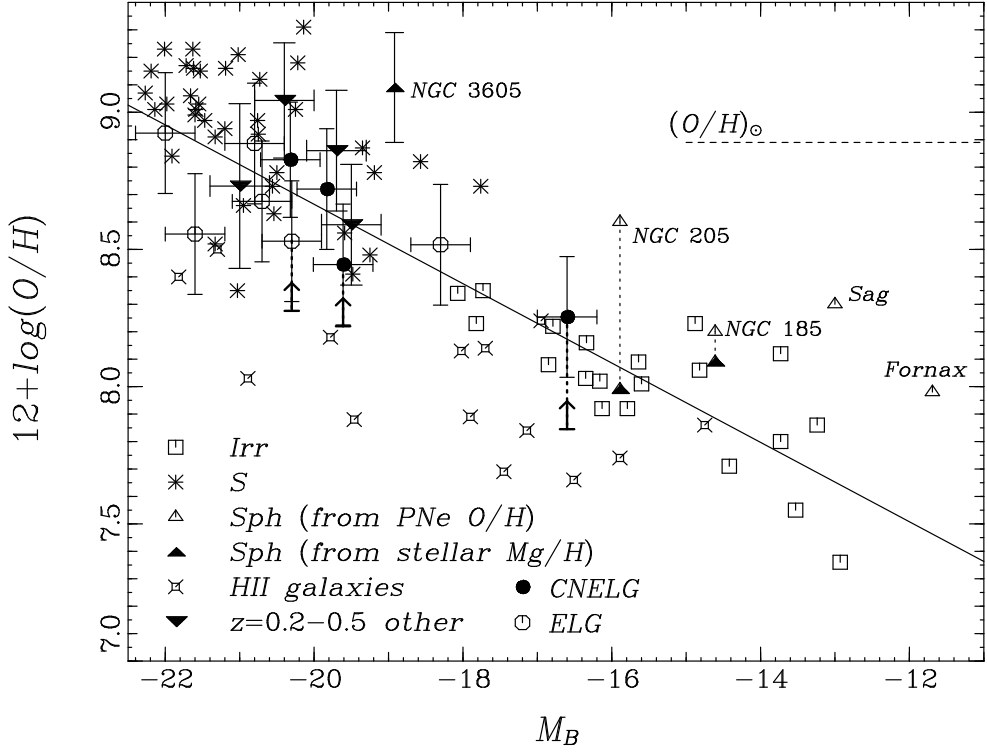}}
\figcaption[M_L.ps] {Oxygen abundance versus absolute blue magnitude
for CNELGs ({\it filled circles}), ELGs ({\it open circles}), and four
field galaxies with emission lines ({\it filled triangles}), nearby dwarf
irregular and blue compact galaxies ({\it open squares;} from Richer \&
McCall 1995), local spiral galaxies ({\it asterisks; ZKH}), local
\HII\ galaxies ({\it stars}), and local dwarf spheroidal galaxies ({\it
open triangles}).  Dotted lines connect oxygen abundance measurements
using empirical methods with lower limits based on direct electron
temperature measurements in three ELGs.  Dotted lines also connect
estimates of the oxygen abundance in NGC~205 and NGC~185 derived from
planetary nebulae (higher values; Richer \& McCall 1995) and stellar
absorption line indices (lower values near 12+log(O/H)=7.89; derived
from Trager \etal\ 1998; see appendix).  The solid line is a least
squares fit to the local irregular and spiral galaxies.  Only
\HII\ galaxies deviate significantly from the local
luminosity-metallicity relation, while the intermediate $z$ ELGs are
consistent with the local relation for Irr and S galaxies.  \label{M_L}}
\end{figure}

\begin{figure}
\centerline{\psfig{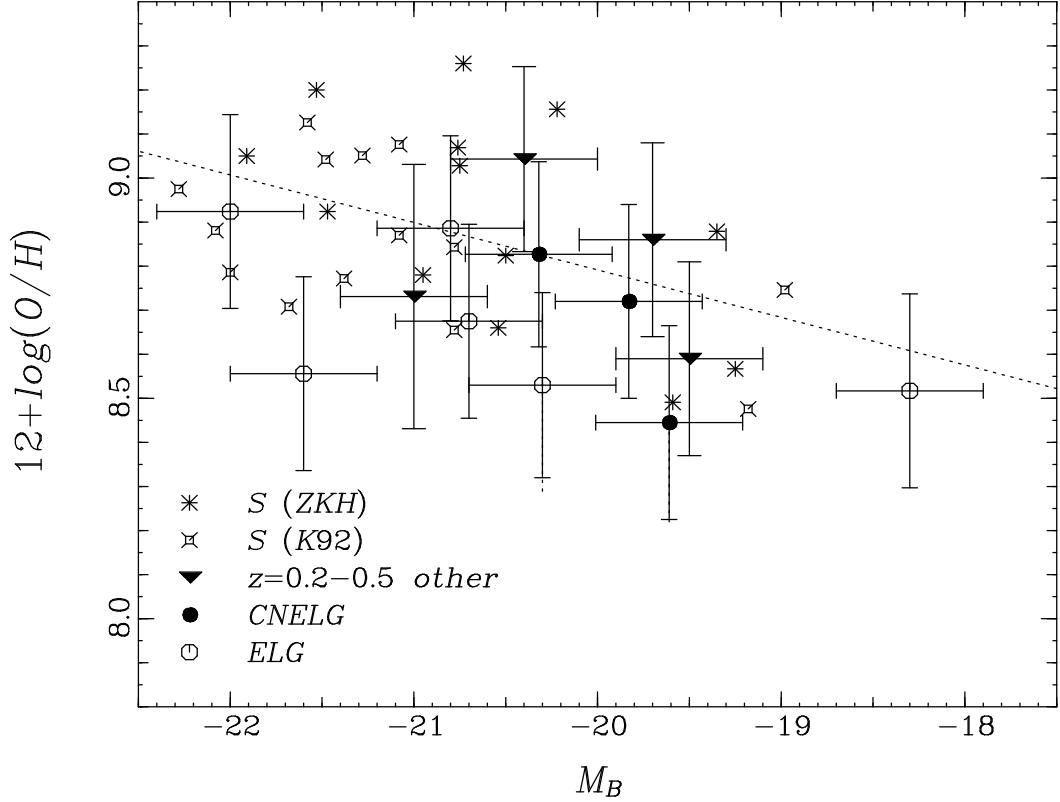}}
\figcaption[M_L3.ps] {The luminosity-metallicity relation 13 objects
from ELG sample, as in previous figures, and for local spirals derived
from global emission line spectra in Kennicutt (1992; 14 galaxies) and
ZKH (13 galaxies).  The dashed line is a least squares fit to the 27
local spiral galaxies.  A linear fit to the entire emission line galaxy
sample is statistically indistinguishable from the spiral data.  The 13
ELGs shown here have lookback times of 3.2---6.4 Gyr.   We find only
marginal evidence for evolution of the luminosity--metallicity relation
based on this small and specifically selected sample of distant
star-forming galaxies.  \label{OH_sigma} }
\end{figure}

\begin{figure}
\centerline{\psfig{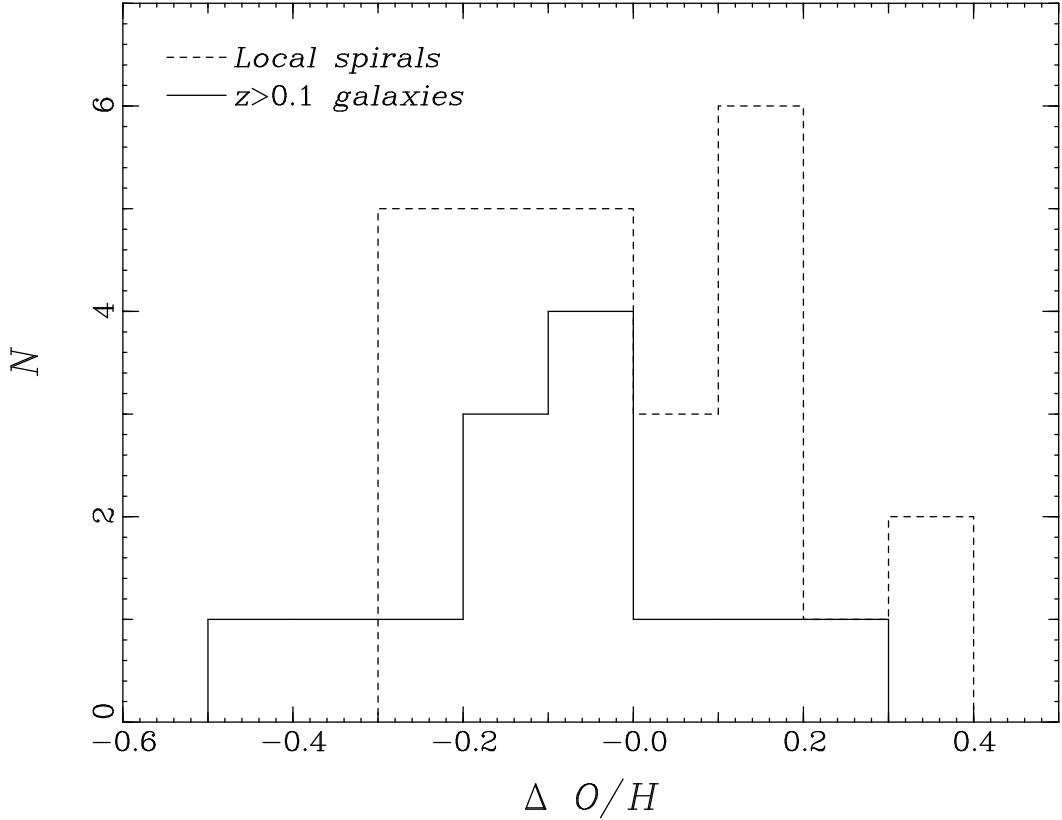}}
\figcaption[OHevol.ps] {Histogram of oxygen abundance residuals,
$\Delta~log(O/H)$, from the best fit linear relationship between
luminosity and oxygen abundance from Figure~6.  The dashed line shows
the residuals for the local spiral galaxies, which are centered about
zero.  The solid line shows the residuals for the 13 galaxies observed
in this work which have a peak at $\Delta~log(O/H)=-0.1$.  For the 13
objects, this constitutes marginal (1.1$\sigma$) evidence that these
$z\simeq$0.4 objects of a given luminosity are slightly oxygen
deficient by $\sim$0.1 dex compared to the present epoch.  However, the
ELGs and CNELGs observed here are selected to be strongly starbursting
objects, so they may not be generally representative of galaxies at
that epoch.  \label{OHevol} }
\end{figure}

\begin{figure}
\centerline{\psfig{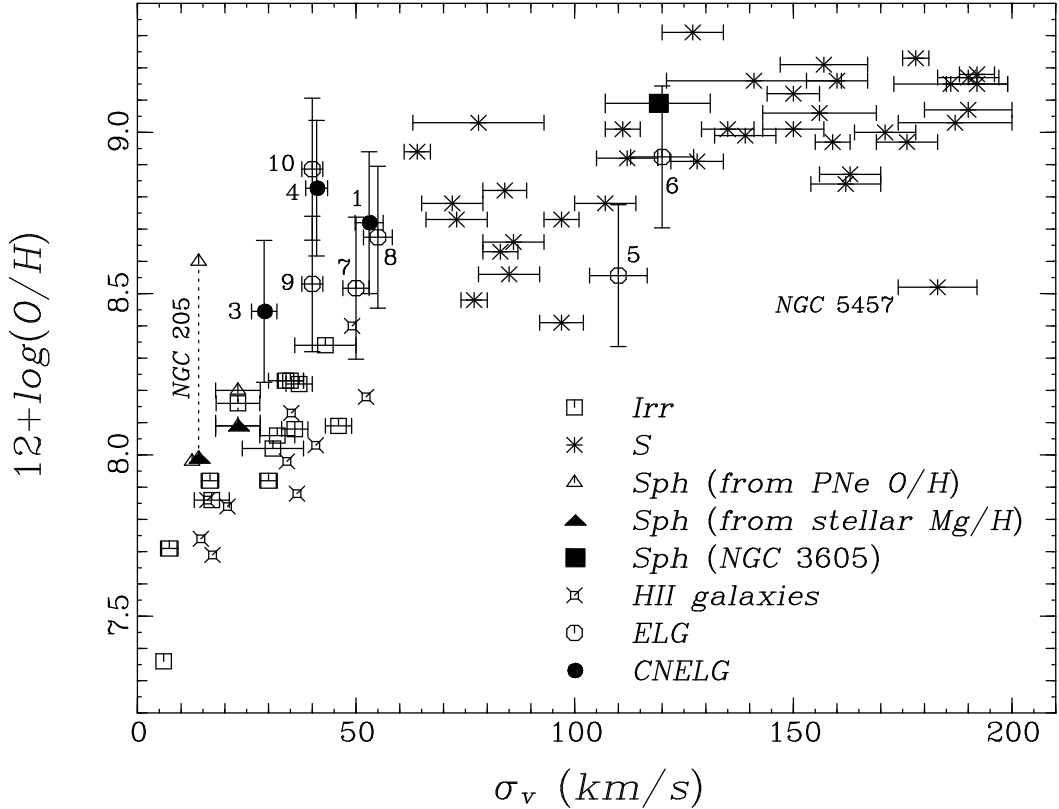}}
\figcaption[OH_sigma.ps] {Oxygen abundance versus velocity dispersion,
for CNELGs ({\it filled circles}),  ELGs ({\it open circles}), nearby
dwarf irregular and blue compact galaxies ({\it open squares}; from
Richer \& McCall 1995), local spiral galaxies ({\it asterisks}; ZKH),
local \HII\ galaxies ({\it stars}), and local dwarf spheroidal galaxies
(open triangles).  The velocity dispersions for spiral and irregular
galaxies are based on \HI\ 21-cm profiles corrected for inclination and
rotation curve shape as in Tully \& Fouque (1985), equation 12.  The
majority of the ELGs cluster to the left of the mean correlation, but
no corrections for the (unknown) inclination or other factors (Rix
\etal\ 1997) which would increase the $\sigma_v$s have been made.
Given a small statistical correction of 50\% for unknown inclinations,
most of the ELGs would lie amidst the massive irregular and low-mas
spiral galaxies.  \label{OH_sigma} }
\end{figure}

\begin{figure}
\centerline{\psfig{file=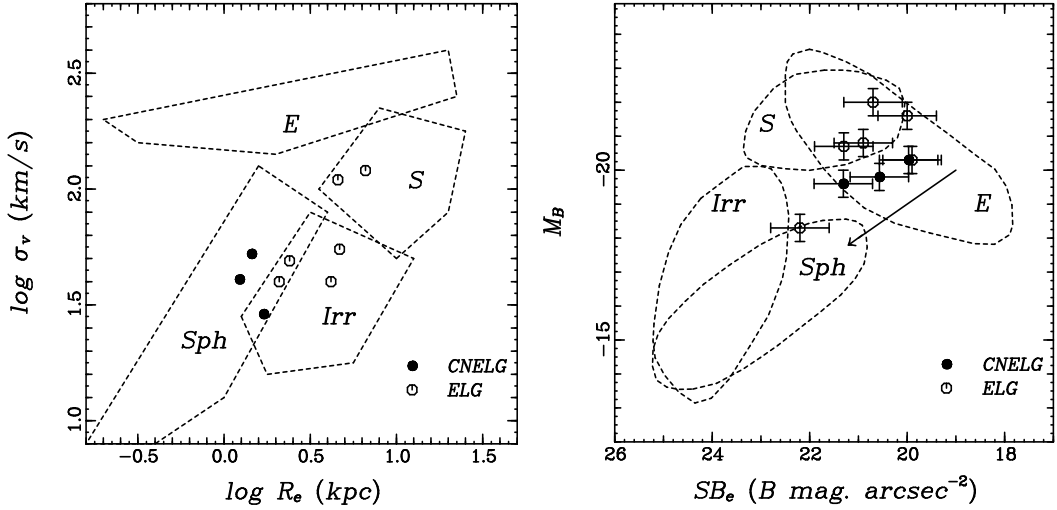,width=5.5in,angle=-90}}
\figcaption[f8.ps] {({\it left panel}): Half light radius versus
velocity width, $\sigma_v$, for CNELGs ({\it filled circles}) and other
emission line galaxies ({\it open circles}). Solid lines illustrate the
locus of local spiral, irregular, elliptical, and spheroidal galaxies,
adapted from Guzman \etal\ (1998).  CNELGs lie in the region occupied
by local spheroidal galaxies suggesting that their descendants will be
related to spheroidals.  The other ELGs are consistent with spiral or
irregular galaxies.  (Right panel):  Mean surface brightness within the
half-light radius versus absolute blue magnitude.  The arrow shows the
direction of anticipated evolution due to fading as a starburst ages.
The high surface brightness of the CNELGs places them within the realm
of elliptical and spheroidal galaxies, while the other ELGs are
consistent with spiral or irregular systems.  \label{SBe_Re} }
\end{figure}

\begin{figure}
\centerline{\psfig{file=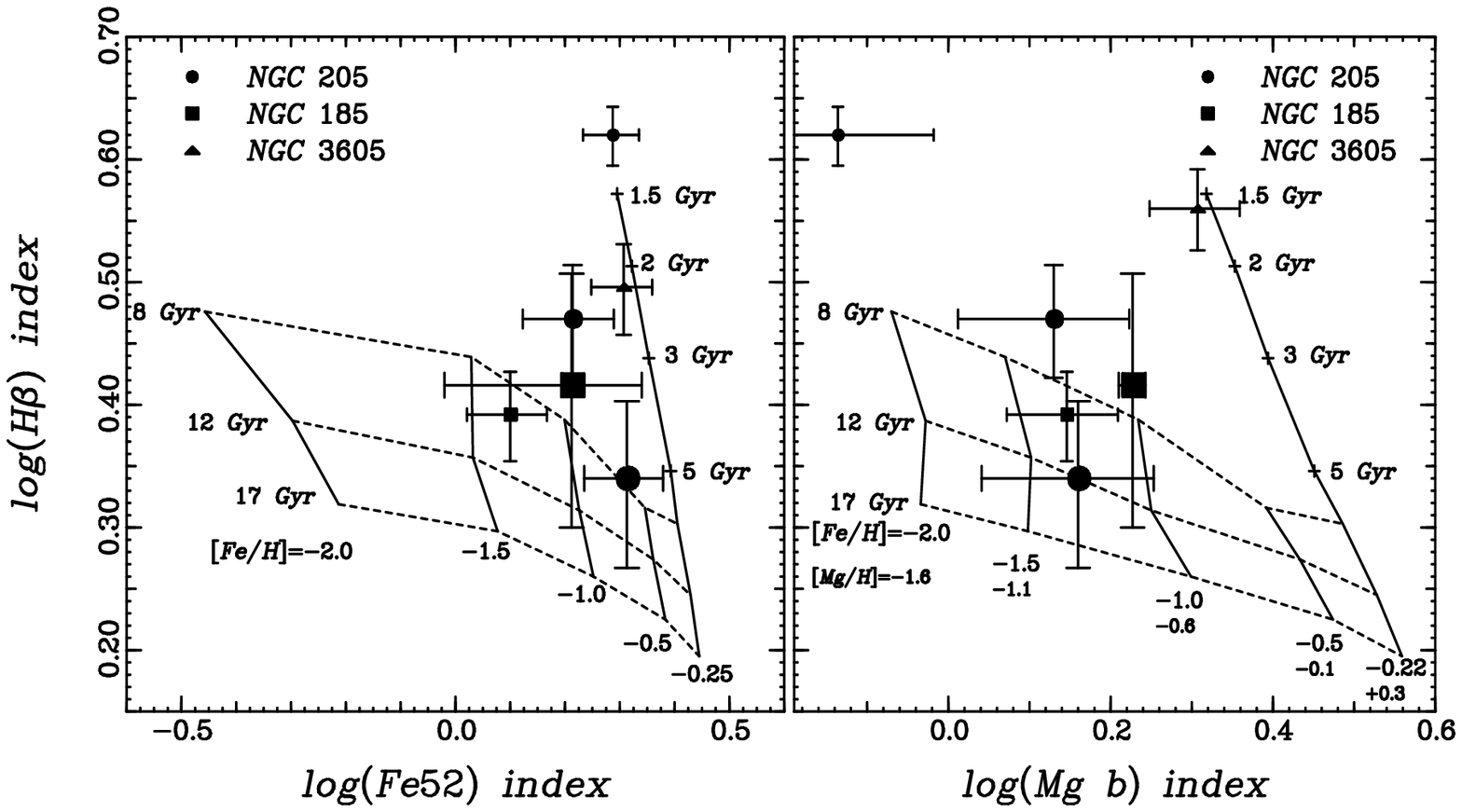,width=5.5in,angle=-90}}
\figcaption[worthey.ps] {Stellar Fe (left) and Mg (right) absorption
line indices versus H$\beta$ index for dwarf spheroidal galaxies
NGC~205, NGC~185, and NGC~3605.  The plot also shows Worthey's (1994)
model grids for single-burst  stellar populations with [Fe/H]$=-2.0$ to
[Fe/H]$=0.25$ and ages from 17 to 8 Gyr.  NGC~205 and NGC~185 data are
generally consistent with $[Fe/H]=-0.7$ to $-1.2$ and $[Mg/H]=-0.8$,
although errors are large, and models for metal--poor populations
younger than 8 Gyr are not available.  \label{worthey} }
\end{figure}


\begin{references}

\reference{} Aller, L.~H. 1942, ApJ, 95, 52
\reference{} Anders, E., \& Grevesse, N. 1989, GeCoA, 53, 197 
\reference{} Bender, R., Burstein, D., \& Faber, S.~M. 1993, ApJ, 399, 462
\reference{} Bender, R., Paquet, A., Nieto, J.~L. 1991, A\&A, 246, 349
\reference{} Bothun, G.~D., Mould, J.~R., Caldwell, N., \& 
 	MacGillivray, H.~T. 1986, ApJ, 311, 526
\reference{} Brodie, J.~P., \& Huchra, J.~P. 1991, ApJ, 379, 157
\reference{} Burstein, D., \& Heiles, C. 1984, ApJS, 54, 33
\reference{} Carney, B. ~W. 1996, PASP, 108, 900
\reference{} Davies, J.~I., \& Phillips, S. 1988, MNRAS, 233, 553
\reference{} Dekel, A., \& Silk, J. 1986, ApJ, 303, 39
\reference{} Dopita, M.~A., \& Evans, I.~N. 1986, ApJ, 307, 431
\reference{} Edmunds, M.~G. \& Pagel, B.~E.~J. 1984, MNRAS, 211, 507
\reference{} Faber, S.~M. 1973, ApJ, 179, 423
\reference{} Filippenko, A.~V. 1982, PASP, 94, 715
\reference{} French, H.~B., 1980, ApJ, 240, 41
\reference{} Fukugita, M., Shimasaku, K., \& Ichikawa, T. 1995, PASP, 107, 945
\reference{} Garnett, D.~R. 1990, ApJ, 363, 142
\reference{} Guzman, R.,Jangren, A., Koo, D.~C., Bershady, M.~A. \&
       Simard, L. 1998, ApJ, 495, L13
\reference{} Guzman, R., Koo, D.~C., Faber, S.~M., Illingworth, G.~D., Takamiya,
 M., Kron, R.~G.,
        \& Bershady, M.~A. 1996, ApJ, 460, L5
\reference{} Howarth, I.~D. 1983, MNRAS, 203, 301
\reference{} Huchtmeier, W.~K., and Richter, O.-G. 1989, A General 
    Catalog of HI Observations of Galaxies, New York:Springer-Verlag
\reference{} Hummer, D.~G., \& Storey, P.~J. 1987, MNRAS, 224, 801
\reference{} Hunter, D.~A. 1994,  AJ, 107, 565
\reference{} Hunter, D.~A., \& Gallagher, J.~S. III. 1997, 475, 65
\reference{} Izotov, Y., Thuan, T.~T., \& Lipovetsky, V.~A. 1994, ApJ, 435, 647
\reference{} Kelson, D. 1998, private comm
\reference{} Kennicutt, R.~C. Jr. 1983, ApJ, 272, 54
\reference{} Kennicutt, R.~C. Jr. 1992, ApJS, 79, 255 (K92)
\reference{} Kobulnicky, H.~A., Kennicutt, R.~C., \& Pizagno, J. 1998,
        ApJ, in prep
\reference{} Kobulnicky, H.~A., \& Skillman, E.~D. 1996, ApJ, 471, 211 
\reference{} Kobulnicky, H.~A., \& Skillman, E.~D. 1997, ApJ, 489, 636
\reference{} Koo, D.~C. 1998, private communication
\reference{} Koo, D.~C., Bershady, M.~A., Wirth, G.~D., Stanford, S.~A.,
        \& Majewski, S.~R. 1994, ApJ, 427, L9
\reference{} Koo, D.~C., Guzman, R., Gallego, J., \& Wirth, G.~D. 1997,
        ApJ, 478, L49
\reference{} Koo, D.~C., Guzman, R., Faber, S.~M., Illingworth, G.~D. Bershady, M.~A., Kron, R.~G.,
	\& Takamiya, M. 1995, ApJ, 440, L49
\reference{} Kormendy, J., \& Bender, R. 1994, in Proc. ESO/OHP Workshop
   on Dwarf Galaxies, ed. G. Meylan \& P. Prugniel (Garching: ESO), 161
\reference{} Kunth, D., \& Sargent, W.~L.~W. 1986, ApJ, 300, 496
\reference{} Labrecque, S., Lucinio, R., Schall, W., Epps H., 
	\& Miller, J. 1995, PASP, 107, 375 
\reference{} Lanzetta, K.~M., Wolf, A., \& Turnshek, D.~A. 1995, ApJ, 440, 433 
\reference{} Larson, R.~B. 1974, MNRAS, 169, 229
\reference{} Lavery, R.~J., Pierce, M.~J., \& McClure, R.~D. 1992, AJ, 104, 2067
\reference{} Lehnert, M.~D., \& Heckman, T.~M. 1996, ApJ, 472, 546
\reference{} Lequeux, J., Peimbert, M., Rayo, J.~F., Serrano, A., \& 
   Torres--Peimbert, S. 1979, A\&A, 80, 155
\reference{} Lilly, S.~J., Le F\'evre, O., Hammer, F., \& Crampton, D. 1996, ApJ, 460, L1
\reference{} Lu, L., Sargent, W.~L.~W., Barlow, T.~.A., Churchill, C.~W., \& Vogt, S. 1997, ApJS, 107,
	475
\reference{} MacLow, M.~M, \& Ferrara, A. 1998, ApJ, in press
\reference{} Madau, P. Ferguson, H.~C., Dickinson, M.~E., Giavalisco, M., Steidel, C.~C., \& Fruchter, A.
	1996, MNRAS, 283, 1388
\reference{} Mateo, M. 1994 in Proc. ESO/OHP Workshop
   on Dwarf Galaxies, ed. G. Meylan \& P. Prugniel (Garching: ESO), 309
\reference{} Mateo, M. 1998, ARA\&A, in press
\reference{} Martin, C.~L. 1998, ApJ, in press
\reference{} Martin, P., \& Roy, J.~R. 1994, 424, 599
\reference{} Marlowe, A.~T., Heckman, T.~M., Wyse, R.~F.~G., \& Schommer,
	R. 1995, ApJ, 438, 563
\reference{} Mathews, W.~G. \& Baker, J.~C. 1971, ApJ, 170, 241
\reference{} McCall, M.~L., Rybski, P.~M., \& Shields, G.~A. 1985, 
	ApJS, 57, 1
\reference{} McWilliam, A., \& Rich, M.~R. 1994, ApJS, 91, 749
\reference{} Munn, J.~A., Koo, D.~C., Kron, R.~G., Majewski, S.~R., 
	Bershady, M~A., Smetanka, J.~J. 1997, ApJS, 109, 45
\reference{} Oke, J.~B. 1990, AJ, 99, 1621
\reference{} Oke, J. B., \etal\ 1995, PASP, 107, 375
\reference{} Osterbrock, D.~E. 1989,  Astrophysics of Gaseous Nebulae
  and Active Galactic Nuclei, University Science Books:Mill Valley CA
\reference{} Pagel, B.~E.~J. 1986, PASP, 98, 1009
\reference{} Pagel, B.~E.~J. Edmunds, M.~G., Blackwell, D.~E., Chun, 
		M.~S., \& Smith, G. 1979, MNRAS, 189, 95
\reference{} Pagel, B.~E.~J., \& Edmunds, M.~G., Fosbury, R.~A.~E., 
	 \& Webster, B.~L. 1978, MNRAS, 184, 569
\reference{} Pagel, B.~E.~J., Edmunds, M.~G., \& Smith, G. 
	1980, MNRAS, 193, 219
\reference{} Pagel, B.~E.~J., Simonson, E.~A., Terlevich, R.~J., \& Edmunds, 
    M.~G. 1992, MNRAS, 255, 325
\reference{} Pagel, B.~E.~J., Terlevich, R.~J., \& Melnick, J.
	 1986, PASP, 98, 1005
\reference{} Peimbert, M. 1967, ApJ, 150, 825 %T fluctuations
\reference{} Peimbert, M. 1975, ARA\&A, 13, 113
\reference{} Peterson, R.~C., Caldwell, N. 1993, AJ, 105, 1411
\reference{} Renzini, A., \& Voli, M. 1981, A\&A, 94, 175 
\reference{} Richer, M.~G., \& McCall, M.~L. 1995, ApJ, 445, 642
\reference{} Rix, H.~W., Guhathakurta, P., Colless, M., \& Ing, K. 1997,
      MNRAS, 285, 779
\reference{} Schlegel, D.~J., Finkbiner, D.~P. \& Davis, M. 1998, preprint
\reference{} Schneider, D.~P., Dressler, A., \& Gunn, J.~E. 1986, AJ, 92, 52
\reference{} Searle, L. 1971, ApJ, 168, 327
\reference{} Searle, L., \& Sargent, W.~L.~W. 1972, ApJ, 173, 25
\reference{} Seaton, M.~J. 1979, MNRAS, 187, 73p
\reference{} Shields, G.~A. 1990, ARA\&A, 28, 525
\reference{} Skillman, E.~D. 1989, ApJ, 347, 883
\reference{} Skillman, E.~D. 1997, Rev. Mex. AASC, 6, 36
\reference{} Skillman, E.~D. 1992, in {\it Elements and the Cosmos}, eds. M.G. Edmunds
and R.J.Terlevich, (Cambridge: University of Cambridge Press) p. 246
\reference{} Skillman, E.~D., \& Kennicutt, R.~C. 1993, ApJ, 411, 655
\reference{} Telles, E., Melnick, J., \& Terlevich, R. 1997, MNRAS, 288, 78
\reference{} Telles, E. \& Terlevich, R. 1997, MNRAS, 286, 183
\reference{} Tenorio-Tagle, G. 1996, AJ, 111, 1641 
\reference{} Terlevich, R. \& Melnick, J. 1981, MNRAS, 195, 839
\reference{} Thuan, T.~X. 1983, ApJ, 268, 667
\reference{} Trager, S. C., Worthey, G., Faber, S. M., Burstein, D., \& Gonzalez, 
	J. J. 1998, ApJS, 116, in press 
\reference{} Tully, R.~B. 1988, { Nearby Galaxies Catalog}, Cambridge
University Press:Cambridge
\reference{} Tully, R.~B., \& Fouque, P. 1985, ApJS, 58, 67
\reference{} van Zee, L., Haynes, M.~P., \& Salzer, J.~J. 1997, AJ, 114, 2479
\reference{} van Zee, L., Skillman, E.~D., \& Salzer, J.~J. 1998, AJ, in press
\reference{} Vila-Costas, M.~B., \& Edmunds, M.~G. 1992, MNRAS, 259, 121
\reference{} Vila-Costas, M.~B., \& Edmunds, M.~G. 1993, MNRAS, 265, 199
\reference{} Walsh, J.~R., Dudziak, G., Minniti, D., \& Zulstra, A.~A. 1997, ApJ, 487, 651
\reference{} Wheeler, J.~C., Sneden, C., \& Truran, J.~W. 1989, ARA\&A,
	27, 279
\reference{} Worthy, G. 1994, ApJS, 95,107
\reference{} Worthy, G. 1998, ApJ, in press
\reference{} Zaritsky, D., Kennicutt, R.~C., \& Huchra, J.~P. 1994, ApJ, 420,
	87 (ZKH)

\end{references}
\end{document}

\begin{planotable}{llccccccccc}
\tablewidth{0pt}
\tablecaption{Log of Keck LRIS Spectroscopy \label{Obs} }
\tablehead{
\colhead{Field} & 
\colhead{Object\tablenotemark{a}} & 
\colhead{RA (2000)}   & 
\colhead{DEC (2000)}   & 
\colhead{z}   & 
\colhead{t$_{lookback}$\tablenotemark{b}} & 
\colhead{$M_B$\tablenotemark{b}} & 
\colhead{$\sigma_v$} & 
\colhead{$R_e$\tablenotemark{c}} & 
\colhead{Refs.} &
\colhead{ID\#}  \\ [.2ex]
\colhead{} &  
\colhead{} &  
\colhead{} &  
\colhead{} &  
\colhead{} &
\colhead{Gyr} &
\colhead{} &
\colhead{(\kms)} &  
\colhead{(kpc)} &  
\colhead{}    &
\colhead{}  } 
\startdata
CNELGs \nl
Lynx-2 &  408115  & 08:43:27.05 & 44:40:09.6 & 0.197 & 3.2 & -19.8 & 53 & 1.44 & 1,2 & 1 \nl
       & 410083   & 08:44:20.98 & 44:45:49.8 & 0.109 & 1.9 & -16.6 & ... & ...   & 3 & 2 \nl
       & 411500   & 08:44:26.26 & 44:49:49.3 & 0.193 & 3.2 & -19.6 & 29 & 1.72 & 1,2 & 3 \nl
                                            \nl
SA-68 & 206134    & 00:18:11.29 & 15:47:51.6 & 0.285 & 4.3 & -20.3 & 41 & 1.22 & 1,2 & 4 \nl
                                            \nl
ELGs \nl
Cl0024+24 & SDG 125 & 00:26:32.04  & 17:10:03.1 & 0.3989 & 5.5 & -21.6 & 110 & 4.6 & 4,5 & 5 \nl
	  & SDG 146 & 00:26:41.92  & 17:09:53.2 & 0.4006 & 5.5 & -22.0 & 120 & 6.7 & 4,5 & 6  \nl
	  & SDG 173A& 00:26:35.99  & 17:09:37.4 & 0.3996 & 5.5 & -18.3 &  50 & 2.4 & 5,6 & 7 \nl
	  & SDG 183 & 00:26:35.99  & 17:09:32.4 & 0.3994 & 5.5 & -20.7 &  55 & 4.7 & 4,5 & 8 \nl
	  & SDG 223 & 00:26:30.84  & 17:09:03.0 & 0.3996 & 5.5 & -20.3 &  40 & 2.1 & 4,5 & 9 \nl
	  & SDG 231 & 00:26:33.87  & 17:08:56.8 & 0.3989 & 5.5 & -20.8 &  40 & 4.2 & 4,5 & 10 \nl
                                                        \nl
Field Objects \nl
Lynx-2    & 408570  & 08:43:31.02 & 44:41:30.3 & 0.496\tablenotemark{h} & 6.4 & -21.0\tablenotemark{d} & ... & ...& 7 & 11 \nl
Lynx-2	  & 408939  & 08:43:34.48 & 44:42:31.4 & 0.425\tablenotemark{h} & 5.7 & -19.5\tablenotemark{e} & ... & ...& 7 & 12 \nl
SA-68     & 206895   & 00:17:44.84 & 15:49:02.5 & 0.4311 & 5.8 & -20.4\tablenotemark{f}               & ... & ...& 8 & 13 \nl
SA-68     & 207213   & 00:17:44.34 & 15:49:34.8 & 0.2343 & 3.7 & -19.7\tablenotemark{g}               & ... & ...& 8 & 14 \nl
\enddata 
\tablerefs{
1.  Guzman \etal\ 1996;
2.  Koo \etal 1994;
3. Guzman, private communication
4.  Schneider, Dressler, \& Gunn 1986;
5.  Koo \etal\ 1997;
6.  Lavery, Pierce, \& McClure 1992;
7. Munn \etal\ unpublished;
8.  Munn \etal\ 1997}
\tablenotetext{a}{Here, object names refer to the full 6-digit nomenclature
of Munn \etal\ (1997).   Koo \etal\ (1995) and Guzman \etal\ (1996) use an abbreviated
4-digit nomenclature for the same objects.}
\tablenotetext{b}{Assumes q$_0$=0.1 and H$_0$=50 \kms\ Mpc$^{-1}$ }
\tablenotetext{c}{Half--light radius}
\tablenotetext{d}{Computed from J=21.6 (Koo 1998, private communication) assuming $q_0$=0.1 and $H_0$=50 \kms\ Mpc$^{-1}$
and a k-correction of $-0.2 mag$ typical for a Scd galaxy at z=0.5 (Fukugita, Shimasaku, \& Ichikawa
1995)}
\tablenotetext{e}{Computed from J=22.7 (Koo 1998, private communication) assuming same cosmology
as above
and a k-correction of $-0.2$ mag typical for a Scd galaxy at z=0.42 (Fukugita, Shimasaku, \& Ichikawa
1995)}
\tablenotetext{f}{Computed from J=21.9 (Munn \etal\ 1997) assuming same cosmology
and a k-correction of $-0.2$ mag typical for a Scd galaxy at z=0.43 (Fukugita, Shimasaku, \& Ichikawa
1995)}
\tablenotetext{g}{Computed from J=21.1 (Munn \etal\ 1997) assuming same cosmology
and a k-correction of $-0.1$ mag typical for a Scd galaxy at z=0.23 (Fukugita, Shimasaku, \& Ichikawa
1995)}
\tablenotetext{h}{Redshifts determined from the data presented
here, and have an RMS accuracy of $\Delta$z=$\pm$0.002}
\end{planotable}
\clearpage

\begin{planotable}{lrrrrrrrr}
\tablecaption{Dereddened Line Strengths for ELGs\tablenotemark{ab}   \label{Craw} }
\tablewidth{0pt}
\tablehead{
\colhead{Line} & 
\colhead{F($\lambda$)} & 
\colhead{L2-408115}   &
\colhead{L2-410083}   & 
\colhead{L2-411500}   & 
\colhead{SA68-206134}  &
\colhead{SDG125}   &
\colhead{SDG146}  &
\colhead{SDG173A}  \\ [.2ex]  
\colhead{}   & 
\colhead{}   & 
\colhead{${I_0}\over{I_{H\beta}}$ }   & 
\colhead{${I_0}\over{I_{H\beta}}$ }   & 
\colhead{${I_0}\over{I_{H\beta}}$ }   & 
\colhead{${I_0}\over{I_{H\beta}}$ }   & 
\colhead{${I_0}\over{I_{H\beta}}$ }   & 
\colhead{${I_0}\over{I_{H\beta}}$ }   & 
\colhead{${I_0}\over{I_{H\beta}}$ }    }
\startdata
3727 [O \II]  &  0.25 & 3.799$\pm$0.367 & 0.601$\pm$0.058 & 3.377$\pm$0.434 & 1.926$\pm$0.106 & 3.672$\pm$0.273 & 1.936$\pm$0.144 & 3.417$\pm$0.258 \nl  
3868 [Ne \III]& 0.23 & $<$0.096        & 0.512$\pm$0.046 & 0.357$\pm$0.042 & 0.127$\pm$0.010 & 0.271$\pm$0.020 & $<$0.070        & $<$0.098        \nl  
4101 H$\delta$& 0.19 & $<$0.219        & 0.261$\pm$0.020 & 0.255$\pm$0.025 & 0.257$\pm$0.043 & 0.273$\pm$0.018 & $<$0.345        & $<$0.130        \nl  
4340 H$\gamma$& 0.12 & 0.430$\pm$0.027 & 0.482$\pm$0.030 & 0.459$\pm$0.035 & 0.436$\pm$0.023 & 0.465$\pm$0.026 & $<$0.363        & 0.484$\pm$0.040 \nl  
4363 [O \III] & 0.12 & $<$0.025        & 0.150$\pm$0.009 & 0.031$\pm$0.008        & $<$0.012        & $<$0.014        & $<$0.051        & $<$0.065        \nl  
4471 He \I    & 0.09 & $<$0.029        & 0.038$\pm$0.003 & $<$0.102        & $<$0.011        & 0.036$\pm$0.005 & $<$0.260        & $<$0.101        \nl  
4686 He \II   & 0.05 & $<$0.024\tablenotemark{c}  & $<$0.008\tablenotemark{c} & 
	0.012$\pm$0.004\tablenotemark{c} & 
	$<$0.011\tablenotemark{c} & $<$0.009\tablenotemark{c} & $<$0.025\tablenotemark{c} & $<$0.090\tablenotemark{c}\nl  
4861 H$\beta$ & 0.00 & 1.000$\pm$0.003 & 1.000$\pm$0.001 & 1.000$\pm$0.003  & 1.000$\pm$0.005 & 1.000$\pm$0.004 & 1.000$\pm$0.010 & 1.000$\pm$0.026 \nl  
4969 [O \III] & -0.02 & 0.424$\pm$0.021 & 2.242$\pm$0.108 & 1.131$\pm$0.055  & 0.669$\pm$0.032 & 0.779$\pm$0.037 & 0.382$\pm$0.020 & 1.032$\pm$0.058 \nl  
5007 [O \III] & -0.04 & 1.264$\pm$0.063 & 6.636$\pm$0.322 & 3.354$\pm$0.168  & 2.091$\pm$0.099 & 2.361$\pm$0.113 & 1.110$\pm$0.054 & 2.798$\pm$0.137 \nl  
5876 He \I    & -0.23 & 0.121$\pm$0.015 & 0.106$\pm$0.010 & 0.092$\pm$0.011  & 0.073$\pm$0.005 & 0.130$\pm$0.010 & $<$0.099        & $<$0.169        \nl  
6312 [S \III] & -0.30 & $<$0.020        & $<$0.005        & $<$0.062         & $<$0.023        & $<$0.003        & $<$0.280        & $<$0.397        \nl  
6548 [N \II]  & -0.34 & 0.271$\pm$0.034 & $<$0.011        &  ...             & $<$0.022        &  ...            & ...             & $<$0.302        \nl  
6564 H$\alpha$& -0.34 & 2.919$\pm$0.363 & 2.862$\pm$0.354 &  ...             & 2.874$\pm$0.176 &  ...            & ...             & $<$0.386        \nl  
6584 [N \II] &  -0.34 & 0.676$\pm$0.085  & 0.024$\pm$0.005 &  ...             & 0.310$\pm$0.020 &  ...            & ...             & $<$0.298        \nl  
6678 He \I   &  -0.35 & $<$0.028         & 0.028$\pm$0.005 &  ...             & 0.113$\pm$0.010 &  ...            & ...             & $<$0.160        \nl  
6717 [S \II] &  -0.36 & 0.608$\pm$0.080  & 0.064$\pm$0.009 &  ...             & 0.406$\pm$0.026 &  ...            & ...             & $<$0.160        \nl
6731 [S \II] &  -0.36 & 0.452$\pm$0.059  & 0.051$\pm$0.007 &  ...             & 0.386$\pm$0.026 &  ...            & ...             & $<$0.160        \nl  
\nl
Abs(H$\beta$)  (\AA)          &  & 2.0            & 2.0           & 2.0           & 8.0           & 2.0           & 2.0       & 2.0\nl  
c(H$\beta$)                   &  & 0.32$\pm$0.05  & 0.04$\pm$0.05 & 0.48$\pm$0.10 & 0.34$\pm$0.05 & 0.85$\pm$0.10 & 0.0$\pm$0.15 & 0.20$\pm$0.10 \nl  
F(H$\beta$) ($\times10^{-16})$  && 2.36$\pm$0.11  & 6.83$\pm$0.32 & 8.12$\pm$0.72 & 2.11$\pm$0.10 & 5.62$\pm$0.26 & 1.96$\pm$0.92& 0.46$\pm$0.03 \nl  
EW(H$\beta$)                   & & 14.0$\pm$0.6   & 211.2$\pm$10.4   & 45.7$\pm$2.3    & 44.8$\pm$2.7    &  46.4$\pm$2.2    & 9.4$\pm$0.5  & 260.5$\pm$15.4    \nl  
n$_e$ (cm$^{-3}$)              & & $\leq$ 100.    & 180$\pm$100.  & $\leq$100\tablenotemark{d} & 450$\pm$300 & $\leq$100\tablenotemark{d} &  $\leq$100\tablenotemark{d} & $\leq$100\tablenotemark{d}  \nl  
\enddata
\tablenotetext{a}{Dereddened relative to H$\beta$ using a Galactic (Seaton 1986) reddening law
			and assuming all of the extinction occurs within the target galaxy.}
\tablenotetext{b}{All upper limits denote 3$\sigma$ upper limits}
\tablenotetext{c}{Includes only narrow, presumably nebular He~II}
\tablenotetext{d}{Assumed value since [S~II] $\lambda\lambda$6717,6731 are not measured}
\end{planotable}
\clearpage

\begin{planotable}{lrrrrrrr}
\tablecaption{Dereddened Line Strengths for ELGs and Field Objects \tablenotemark{ab}   \label{Craw} }
\tablewidth{0pt}
\tablehead{
\colhead{Line} & 
\colhead{SDG183}   &
\colhead{SDG223}   & 
\colhead{SDG231}   & 
\colhead{L2-408570}  &
\colhead{L2-408939}   &
\colhead{SA68-206895}  &
\colhead{SA68-207213}  \\ [.2ex]  
\colhead{}   & 
\colhead{${I_0}\over{I_{H\beta}}$ }   & 
\colhead{${I_0}\over{I_{H\beta}}$ }   & 
\colhead{${I_0}\over{I_{H\beta}}$ }   & 
\colhead{${I_0}\over{I_{H\beta}}$ }   & 
\colhead{${I_0}\over{I_{H\beta}}$ }   & 
\colhead{${I_0}\over{I_{H\beta}}$ }   & 
\colhead{${I_0}\over{I_{H\beta}}$ }    }
\startdata
3727 [O \II]  & 2.984$\pm$0.223 & 1.779$\pm$0.133 & 3.534$\pm$0.266 & 3.408$\pm$0.747 & 3.028$\pm$0.226 & 1.971$\pm$0.247 & 2.541$\pm$0.328 \nl  
3868 [Ne \III]& $<$0.059        & 0.243$\pm$0.017 & $<$0.155        & $<$0.144        & $<$0.085        & $<$0.050        & $<$0.090        \nl  
4101 H$\delta$& $<$0.247        & 0.278$\pm$0.018 & $<$0.363        & $<$0.346        & $<$0.297        & $<$0.289        & $<$0.188        \nl  
4340 H$\gamma$& 0.411$\pm$0.036 & 0.460$\pm$0.026 & 0.443$\pm$0.041 & $<$0.304        & 0.471$\pm$0.035 & $<$0.316        & 0.727$\pm$0.062 \nl  
4363 [O \III] & $<$0.033        & 0.027$\pm$0.002 & $<$0.015        & $<$0.120        & $<$0.078        & $<$0.095        & $<$0.084        \nl  
4471 He \I    & $<$0.037        & 0.035$\pm$0.004 & $<$0.014        & $<$0.090        & $<$0.069        & $<$0.069        & $<$0.078        \nl  
4686 He \II   & $<$0.036\tablenotemark{c}  & $<$0.009\tablenotemark{c} & 
	$<$0.045\tablenotemark{c} & 
	$<$0.090\tablenotemark{c} & $<$0.055\tablenotemark{c} & $<$0.061\tablenotemark{c} & $<$0.077\tablenotemark{c}\nl  
4861 H$\beta$ & 1.000$\pm$0.018 & 1.000$\pm$0.003 & 1.000$\pm$0.011  & 1.000$\pm$0.059 & 1.000$\pm$0.024 & 1.000$\pm$0.026 & 1.000$\pm$0.027 \nl  
4969 [O \III] & 0.601$\pm$0.030 & 1.292$\pm$0.061 & 0.169$\pm$0.018  & 0.627$\pm$0.056 & 0.943$\pm$0.050 & 0.250$\pm$0.022 & 0.471$\pm$0.034 \nl  
5007 [O \III] & 2.010$\pm$0.097 & 4.050$\pm$0.194 & 0.438$\pm$0.026  & 1.422$\pm$0.081 & 2.608$\pm$0.126 & 0.602$\pm$0.047 & 1.364$\pm$0.074 \nl  
5876 He \I    & $<$0.043        & 0.116$\pm$0.009 & $<$0.059         & ...             & $<$0.271        & $<$0.933        & $<$0.121        \nl  
6312 [S \III] & $<$0.461        & $<$0.041        & $<$0.189         & ...             & $<$0.058        & $<$0.208        & $<$0.133        \nl  
6548 [N \II]  & $<$0.204        & 0.045$\pm$0.010 & $<$0.169         & ...             & ...             & ...             & $<$0.150        \nl  
6564 H$\alpha$& 2.476$\pm$0.236 & 2.904$\pm$0.264 & 2.856$\pm$0.267  & ...             & ...             & ...             & 2.089$\pm$0.358 \nl  
6584 [N \II] & $<$0.204         & 0.245$\pm$0.024 & $<$0.164         & ...             & ...             & ...             & 0.406$\pm$0.065 \nl  
6678 He \I   & $<$0.212         & $<$0.028        & $<$0.154         & ...             & ...             & ...             & $<$0.112        \nl  
6717 [S \II] & $<$0.033         & $<$0.093        & $<$0.162         & ...             & ...             & ...             & 0.830$\pm$0.153   \nl
6731 [S \II] & $<$0.033         & $<$0.093        & $<$0.162         & ...             & ...             & ...             & 0.755$\pm$0.141        \nl  
\nl
Abs(H$\beta$)  (\AA)          & 4.0            & 8.0            & 3.0           & 2.0           & 2.0           & 2.0       & 2.0\nl  
c(H$\beta$)                   & 0.00$\pm$0.10  & 0.07$\pm$0.05  & 0.45$\pm$0.05 & 0.00$\pm$0.15 & 0.20$\pm$0.10 & 0.00$\pm$0.15 & 0.00$\pm$0.15 \nl  
F(H$\beta$) ($\times10^{-16})$& 1.52$\pm$0.07  & 3.48$\pm$0.16  & 1.15$\pm$0.06 & 2.11$\pm$0.10 & 0.73$\pm$0.04 & 0.53$\pm$0.03& 0.61$\pm$0.03 \nl  
EW(H$\beta$)                  & 23.5$\pm$1.2   & 140.5$\pm$6.1     & 15.1$\pm$0.7  & 44.8$\pm$2.6    &  34.3$\pm$2.2    & 8.7$\pm$0.5  & 9.4$\pm$0.5    \nl  
n$_e$ (cm$^{-3}$)             & $\leq$ 100.\tablenotemark{d}    & $\leq$ 100.\tablenotemark{d}  & $\leq$100.\tablenotemark{d} & $\leq$100.\tablenotemark{d} &  $\leq$100\tablenotemark{d} & $\leq$100\tablenotemark{d} & 550$\pm$300   \nl  
\enddata
\tablenotetext{a}{Dereddened relative to H$\beta$ using a Galactic (Seaton 1986) reddening law
			and assuming all of the extinction occurs within the target galaxy.}
\tablenotetext{b}{All upper limits denote 3$\sigma$ upper limits}
\tablenotetext{c}{Includes only narrow, presumably nebular He~II}
\tablenotetext{d}{Assumed value since [S~II] $\lambda\lambda$6717,6731 are not measured}
\end{planotable}
\clearpage

\begin{planotable}{lrrrrrr}
\tablecaption{Derived Parameters and Abundances   \label{phys} }
\tablewidth{0pt}
\tablehead{
\colhead{}   &
\colhead{$T(O^{++})$}   &
\colhead{$T(O^{+})$}   & 
\colhead{12+log(O/H) }   & 
\colhead{$T(O^{++})$}  &
\colhead{12+log(O/H)\tablenotemark{a}}   &
\colhead{log(N/O)}  \\ [.2ex]  
\colhead{ Object }   & 
\colhead{ (K) }   & 
\colhead{ (K) }   & 
\colhead{ direct}   & 
\colhead{ indicative}   & 
\colhead{empirical\tablenotemark{a}}   & 
\colhead{}    }
\startdata
L2-408115      & $<$15100       & $<$13700        & $>$7.8                 & 6700$\pm$350  & 8.72$\pm$0.07 & -1.17$^{+0.12}_{-0.16}$                     \nl  
L2-410083     & 16100$\pm$ 500 & 14000$\pm$ 700  & 7.84$^{+0.03}_{-0.03}$ & 11300$\pm$ 500& 8.25$\pm$0.09 & -1.59$^{+0.07}_{-0.08}$                    \nl  
L2-411500     & 11000$\pm$ 500 & 11700$\pm$ 700  & 8.22$^{+0.04}_{-0.04}$ & 11100$\pm$500 &  8.44$\pm$0.09&  ...                                      \nl  
SA68-206134      & $<$9500        & $<$10800        & $>$8.2                 & 6400$\pm$250  & 8.83$\pm$0.05 & -1.23$^{+0.12}_{-0.16}$                   \nl  
SDG125       & $<$12300       & $<$12400        & $>$8.0                 & 7900$\pm$350  & 8.57$\pm$0.09 & ...                                      \nl  
SDG146       & $<$24000       & $<$16200        & $>$7.3                 & 5700$\pm$250  & 8.92$\pm$0.06 & $<$-0.88                                 \nl  
SDG173A      & $<$18300       & $<$14900        & $>$7.7                 & 8400$\pm$800  & 8.52$\pm$0.09 & $<$-1.33                    \nl  
SDG183       & $<$14400       & $<$13400        & $>$7.3                 & 6700$\pm$500  & 8.68$\pm$0.09 & -1.08$^{+0.14}_{-0.21}$                   \nl
SDG223       & 10100$\pm$ 390 & 11200$\pm$ 600  & 8.28$^{+0.03}_{-0.04}$ & 8500$\pm$300  & 8.53$\pm$0.08 & -1.15$^{+0.10}_{-0.12}$                    \nl
SDG231       & $<$19400       & $<$15200        & $>$7.5                 & 5400$\pm$250  & 8.89$\pm$0.09 & -1.36$^{+0.15}_{-0.23}$                    \nl
L2-408570      & $<$40000       & $<$19000        & $>$7.2                 & 6600$\pm$1000 & 8.73$\pm$0.25 & ...                                       \nl
L2-408939      & $<$18500       & $<$14900        & $>$7.7                 & 7800$\pm$700  & 8.60$\pm$0.08 & ...                                       \nl
SA68-206895      & $<$40000       & $<$20000        & $>$7.0                 & 5000$\pm$200  & 9.04$\pm$0.06 & ...                                     \nl
SA68-207213      & $<$31000       & $<$18000        & $>$7.2                 & 6000$\pm$300  & 8.86$\pm$0.07 & -1.27$^{+0.14}_{-0.21}$             \nl
\enddata
\tablenotetext{a}{The listed uncertainties include 
measurement uncertainties only, based on
errors in the tabulated line strengths. 
Systematic errors in the $R_{23}$ strong-line relation which is 
calibrated from photoionization models 
may run as high as 0.2 dex.  Elsewhere in this paper, an error of 0.2 dex has been
 added in quadrature with the uncertainty listed here to obtain a total uncertainty on O/H.}
\end{planotable}
\clearpage

\begin{planotable}{lccccc}
\tablecaption{Other Derived Parameters  \label{phys} }
\tablewidth{0pt}
\tablehead{
\colhead{}   &
\colhead{$L_{H\beta}$\tablenotemark{a}}   &
\colhead{$L_{H\beta}$\tablenotemark{b}}   &
\colhead{$Q_{Ly}$\tablenotemark{c}} & 
\colhead{SFR\tablenotemark{d}}  &
\colhead{$M_{H^+, He^+}$\tablenotemark{e}}  \\ [.2ex] 
\colhead{ Object }   & 
\colhead{$\times10^{41} erg~s^{-1}$}   & 
\colhead{$\times10^{41} erg~s^{-1}$}   & 
\colhead{$10^{53}~s^{-1}$}    &
\colhead{$M_\odot~yr^{-1}$}  &
\colhead{$10^6 M_\odot$}   } 
\startdata
L2-408115    & 0.46    &  0.96 & 2.02  & 2.43  & 8.47   \nl  
L2-410083   & 0.38    &  0.42 & 0.84  & 1.05  & 3.52   \nl  
L2-411500   & 1.53    &  4.62 & 9.64  & 11.90 & 40.4   \nl  
SA68-206134    & 0.94    &  2.06 & 4.30  & 5.26  & 18.2   \nl  
SDG125     & 5.04    &  39.7 & 82.1  & 100.95 & 344 \nl  
SDG146     & 1.75    &  1.96 & 4.09  & 5.05 & 17.1 \nl  
SDG173A    & 0.41    &  0.73 & 1.52  & 1.85 & 6.38 \nl  
SDG183     & 1.36    &  1.52 & 3.18  & 3.86 & 13.3 \nl
SDG223     & 3.13    &  3.09 & 6.48 & 7.86 & 27.2 \nl
SDG231     & 1.03    &  3.24 & 6.77 & 8.26 & 28.4 \nl
L2-408570    & 3.29    &  2.11 & 4.41  & 5.38 & 18.5 \nl
L2-408939    & 0.80    &  1.16 & 2.42  & 2.97 & 10.2 \nl
SA68-206895    & 0.49&  0.52 & 1.09  & 1.31 & 4.57 \nl
SA68-207213    & 0.18&  0.61 & 1.27  & 1.54 & 5.32 \nl
\enddata
\tablenotetext{a}{Rest-frame H$\beta$ luminosity computed using the
  Redshifts from Table~1, H$\beta$ fluxes from Tables~2---3 assuming $H_0$=50 \kms\ Mpc$^{-1}$ and
  $q_0$=0.1. } 
\tablenotetext{b}{Corrected for extinction at H$\beta$ using the
  extinction estimate from Tables~2-3.}
\tablenotetext{c}{Assumes Case B recombination, that
  all of the ionizing photons and Lyman series photons are re-processed 
  within the galaxy and emerge as Balmer
  series transitions.  Explcitly, $Q_{Ly}=
  {{1}\over{h\nu_{H\beta}}}{{\alpha_B(10000 K)}\over{\alpha_{H\beta}^{eff}(10000 K)}} 
	L_{H\beta} =2.09\times10^{12}~L_{H\beta}~(erg~s^{-1})$. This should 
   be regarded as a lower limit on $Q_{Ly}$
   since the galaxies may be translucent to Lyman continuum photons.}
\tablenotetext{d}{Computed assuming $L_{H\alpha}/L_{H\beta}=2.86$ and
  $SFR (M_\odot~yr^{-1}) = L_{H\alpha}/1.12\times10^{41}$ as in Kennicutt
  (1983). }
\tablenotetext{e}{Ionized gas mass, using the H$\beta$ luminosity from
  column 1, assuming recombination coefficients for H$^+$ 
  at $T_e$=10,000 K and n$_e=100~cm^{-3}$ (Osterbrock 1989), multiplied
  by a factor of 1.3 to account for ionized He.    }
\end{planotable}